\documentclass[a4paper,english]{article}
\usepackage[margin=3.25cm]{geometry}

\usepackage[utf8]{inputenc}
\usepackage{babel}
\usepackage{stmaryrd}
\usepackage{latexsym}
\usepackage{amsmath, amsthm, amssymb} 
\usepackage{mathtools}
\usepackage{logic-thm}
\usepackage{xcolor}
\usepackage{graphicx}
\usepackage{wrapfig}
\usepackage[textsize=scriptsize,disable]{todonotes}
\usepackage[textsize=scriptsize]{todonotes}
\usepackage{comment}
\usepackage{tikz}

\usepackage{amsmath}
\newtheorem{theorem}{Theorem}
\newtheorem{proposition}[theorem]{Proposition}
\newtheorem{corollary}[theorem]{Corollary}
\newtheorem{lemma}[theorem]{Lemma}
\newtheorem{definition}[theorem]{Definition}
\newtheorem{example}[theorem]{Example}

\makeatletter
\newtheorem*{rep@theorem}{\rep@title}
\newcommand{\newreptheorem}[2]{
  \newenvironment{rep#1}[1]{\medskip\def\rep@title{\bfseries #2 \ref{##1}}                            \begin{rep@theorem}}                           {\end{rep@theorem}}}
\newreptheorem{theorem}{Theorem}
\newreptheorem{lemma}{Lemma}
\newreptheorem{corollary}{Corollary}
\newreptheorem{definition}{Definition}
\newreptheorem{proposition}{Proposition}
\newreptheorem{example}{Example}
\newreptheorem{remark}{Remark}
\makeatother

\renewcommand\paragraph[1]{\par\smallskip\noindent{\bfseries #1~}}

\usetikzlibrary{automata,arrows,decorations.pathreplacing,patterns}
\tikzstyle{hidden}=[minimum size=1mm, shape=circle, inner sep=0pt, outer sep=0pt, style={font=\vphantom{Ag}}]
\tikzstyle{dot}=[draw,shape=circle, fill, minimum size=1mm, inner sep=0pt, outer sep=0pt]
\tikzstyle{state}=[draw,shape=circle, fill=gray!25, minimum size=6mm, inner sep=0pt, outer sep=0pt, style={font=\vphantom{Ag}}]
\tikzstyle{letter}=[draw,shape=rectangle, fill=gray!25, minimum size=5mm, inner sep=0pt, outer sep=0pt, style={font=\vphantom{Ag}}]
\tikzstyle{arrow}=[->, >=stealth', shorten >=1pt]

\newsavebox{\wraptext}
\newsavebox{\wrapfig}
\newsavebox{\deftext}
\savebox{\deftext}{Xp}
\newcommand\mywraptextjustified[2]{\savebox{\wraptext}{\parbox[t]{\linewidth-#1}                                                                 {#2\linebreak}}}

\newcommand\mywrapfig[2]{\savebox{\wrapfig}{\parbox[t]{#1}{\centering #2}}                         \usebox{\wraptext}                         \raisebox{-\ht\wrapfig}{\usebox{\wrapfig}}                         \vspace{-\ht\deftext minus 1pt}\linebreak}

 \DeclareFontFamily{U}{mathb}{\hyphenchar\font45}
\DeclareFontShape{U}{mathb}{m}{n}{
  <-6> mathb5 <6-7> mathb6 <7-8> mathb7
  <8-9> mathb8 <9-10> mathb9
  <10-12> mathb10 <12-> mathb12
}{}
\DeclareSymbolFont{mathb}{U}{mathb}{m}{n}
\DeclareMathSymbol{\righttoleftarrow}{\mathrel}{mathb}{"FD}
\newcommand{\lefttorightarrow}{{\mathpalette\mirrorrighttoleftarrow\relax}}
\newcommand{\mirrorrighttoleftarrow}[2]{\reflectbox{$#1{\mkern3mu\righttoleftarrow}$}}
\newcommand{\mycomment}[1]{}

\renewcommand{\act}[2][]{\mathrel{\raisebox{-1pt}[10pt][0pt]{  \ensuremath{\underset{^{\raisebox{-6pt}[0pt][0pt]{\ensuremath{^{^{#1}}}}}}                       {\raisebox{0pt}[3pt][0pt]{\ensuremath{\relbar\mspace{-8mu}\xrightarrow{#2}}}}}}}}

\newcommand{\cupdot}{\mathop{\mathaccent\cdot\cup}}

\newcommand{\shortcut}{\mathsf{Shortcut}}
\newcommand{\norm}{\mathsf{Norm}}
\newcommand{\busy}{\mathsf{Busy}}

\newcommand{\Qright}{Q_{\succ}}
\newcommand{\Qleft}{Q_{\prec}}

\newcommand{\Qlefta}{Q_{1,\prec}}

\newcommand{\Qleftb}{Q_{2,\prec}}
\newcommand{\lft}{\mathsf{left}}
\newcommand{\rgt}{\mathsf{right}}

\newcommand{\lftmark}{\mathop{\vdash}}
\newcommand{\rgtmark}{\mathop{\dashv}}
\newcommand{\NFT}{{2NFT}\xspace}

\newcommand{\NFA}{{2NFA}\xspace}

\newcommand\bin[2]{\big(\begin{smallmatrix}#1\\#2\end{smallmatrix}\big)}
\newcommand\binleadsto{\mathrel{\begin{smallmatrix}\leadsto\\\leadsto\end{smallmatrix}}}
\newcommand\oelement{\sigma}
\newcommand\In{\mathsf{in}}
\newcommand\Out{\mathsf{out}}
\newcommand\Orig{\mathsf{orig}}
\newcommand\prj[2][]{\mathord{\downharpoonright_{#2}^{#1}}}
\newcommand\lift[2][]{\mathord{\upharpoonright_{#1}^{#2}}}
\newcommand\otype{\tau}
\newcommand\bbB{\mathbb{B}}

\renewcommand\true{\mathit{true}}

\title{\bfseries Origin-equivalence of two-way word transducers is in PSPACE}
\author{Sougata Bose\\ LaBRI, University of Bordeaux \and
        Anca Muscholl\\ LaBRI, University of Bordeaux \and
        Vincent Penelle\\ LaBRI, University of Bordeaux \and
        Gabriele Puppis\\ \phantom{xxxxxi}CNRS, LaBRI\phantom{xxxxxi}}
\date{}

\begin{document}

\maketitle
\setlength{\marginparwidth}{35mm}
\setlength{\marginparsep}{1mm}

\begin{abstract}
We consider equivalence and containment problems for word transductions.
These problems are known to be undecidable when the transductions
are relations between words realized by non-deterministic transducers, 
and become decidable when restricting to functions from words to words. 
Here we prove that decidability can be equally recovered the
\emph{origin semantics}, that was introduced by Boja\'nczyk in 2014. We prove that the equivalence
and containment problems for two-way word transducers in the origin semantics
are \PSPACE-complete.
We also consider a variant of the containment problem where two-way 
transducers are compared under the origin semantics, but in a more
relaxed way, by allowing distortions of the origins.
The possible distortions are described by means of a \emph{resynchronization}
relation. We propose MSO-definable resynchronizers and show that they preserve the decidability of
the containment problem under resynchronizations.
\end{abstract}

\section{Introduction}

Finite-state transducers over words were studied in computer science
very early, at the same time as finite-state automata, see
e.g.~\cite{sch61,ahu69,eilenberg1974automata,ber79}. A transducer 
defines a binary relation between words by associating an output 
with each transition. It is called functional if this relation is 
a partial function. Whereas the class of functions defined by one-way 
transducers, i.e.~transducers that process their input from 
left to right, has been extensively considered in the past, 
the study of two-way transducers is quite recent. 
Connections to logic, notably through the notion of graph
transformations definable in monadic second-order logic
(MSO)~\cite{Cou97,eng97}, have shown that the functions
realized by two-way word transducers can be equally defined
in terms of MSO transductions~\cite{eh01}. 
This result is reminiscent of Büchi-Elgot characterizations 
that hold for many classes of objects (words, trees, traces, etc). 
For this reason, transductions described by functional two-way 
word transducers are called \emph{regular functions}. 
For a recent, nice  survey on logical and algebraic
properties of regular word transductions the
reader is referred to~\cite{FR16siglog}.

Non-determinism is a very natural and desirable feature for most types
of automata. However, for word transducers, non-determinism means less
robustness. As an example, non-deterministic transducers are not
equivalent anymore to NMSOT (the non-deterministic version of MSO
transductions), however the latter is  equivalent to non-deterministic
streaming transducers~\cite{ad11}. A major problem is the
undecidability of the equivalence of non-deterministic, one-way word
transducers~\cite{gri68} (also called NGSM, and
capturing the class of \emph{rational relations}). In contrast,
equivalence of functional, two-way word transducers is
decidable~\cite{culik-karhumaki87}, even in \PSPACE. 
This complexity is mainly based on the fact that it can be checked in \PSPACE 
if two non-deterministic, two-way automata are equivalent (see
e.g.~\cite{vardi89twoway}). 

The equivalence test is one of the most widely used operation on
automata, so that it becomes a natural question to know what is needed
to recover decidability of equivalence for rational
relations, and even for \emph{regular relations}, which are transductions 
defined by non-deterministic, two-way transducers. The main result of our paper
is that equivalence of non-deterministic, two-way transducers is decidable 
if one adopts a semantics based on origin information. 
According to this \emph{origin semantics}~\cite{boj14icalp}, each letter of
the output is tagged with the input position that generated it. Thus, a 
relation in the origin semantics becomes a relation over
$\S^* \times (\G \times \Nat)^*$. Surprisingly, the complexity of the
equivalence test turns out to be as low as it could be, namely in
\PSPACE (thus \PSPACE-complete for obvious reasons).

As a second result, we introduce a class of MSO-definable
resynchronizations for regular relations. A resynchronization~\cite{FiliotJLW16}
is a binary relation over $\S^* \times (\G \times \Nat)^*$ that preserves
the input and the output (i.e., the fist two components), but can
change the origins (i.e., the third
component). Resynchronizations 
allow to compare transducers under the origin semantics in a more
relaxed way, by allowing distortions of origins. Formally,
given two non-deterministic, two-way transducers $\Tt_1,\Tt_2$, and a
resynchronization $R$, we want to compare $\Tt_1,\Tt_2$ under the
origin semantics modulo $R$. Containment of $\Tt_1$ in $\Tt_2$ modulo
$R$ means that for each tagged input/output pair $\s'$ generated by
$\Tt_1$, there should be some tagged input/output pair $\s$ generated
by $\Tt_2$ such that $(\s,\s') \in R$. In other words, the
resynchronization $R$ describes possible distortions of origin,
and we ask whether $\Tt_1$ is contained in $R(\Tt_2)$. The
resynchronizations defined here correspond to MSO formulas that describe the change
of origin by mainly considering the input (and to some small extent,
the output). The containment problem under such resynchronizations
turns to be undecidable, unless we enforce some restrictions. It is
decidable for those resynchronizations $R$ that use formulas satisfying 
a certain (decidable) `boundedness' property. In addition, if $R$ is fixed,
then the containment problem modulo $R$ is solvable in \PSPACE,
thus with the same complexity as the origin-equivalence problem. 
This is shown by providing a two-way transducer $\Tt'_2$ that is 
equivalent to $R(\Tt_2)$
(we say that $R(\Tt_2)$ is realizable by a transducer), 
and then we check containment of $\Tt_1$ in $\Tt'_2$ using
our first algorithm. We conjecture that our class of
resynchronizations captures the rational resynchronizations
of~\cite{FiliotJLW16}, but we leave this for future work.

\paragraph{Related work and discussion.} 
The origin semantics for transducers has been
introduced in~\cite{boj14icalp}, and was shown to enjoy several nice
properties, in particular a Myhill-Nerode characterization that can be
used to decide the membership problem for subclasses of transductions,
like first-order definable ones. The current state of the art 
counts quite a number of results related to the origin semantics 
of transducers. In~\cite{BDGPicalp17} a characterization of the class 
of origin graphs generated by (functional) streaming transducers
is given (the latter models were studied in~\cite{AlurCerny10,ad11}). 
Decision problems for tree transducers under the origin semantics have been
considered in~\cite{fmrt18IC}, where it is shown that 
origin-equivalence of top-down tree transducers is decidable. Note
that top-down transducers correspond on words to one-way transducers,
so the result of~\cite{fmrt18IC} is incomparable with ours.

As mentioned before, the idea of resynchronizing origins of word transducers has been
introduced by Filiot et al.~in 
\cite{FiliotJLW16} for the case of one-way transducers. Rational resynchronizers
as defined in~\cite{FiliotJLW16} are one-way transducers $\Rr$ that read
sequences of the form $u_1v_1u_2v_2 \cdots u_nv_n$, where $u_1 \cdots
u_n$ represents the input and  $v_1 \cdots v_n$ the output, with the
origin of $v_i$ being the last letter of $u_i$. It is required that
any image of $u_1v_1u_2v_2 \cdots u_nv_n$ through $\Rr$ leaves the
input and the output part unchanged, thus only  origins change. The
definition of resynchronizer in the one-way case is natural. However,
it cannot be extended to two-way transducers, since there is no 
word encoding of tagged input-output pairs realized by arbitrary two-way transducers. \
Our approach is logic-based: we define MSO resynchronizations that
refer to origin graphs. More precisely, our  MSO resynchronizations
are formulas that talk about the input and, to a limited extent, about
the tagged output.

\paragraph{Overview.} 
After introducing the basic definitions and notations
in Section~\ref{sec:preliminaries}, we present the main result about
the equivalence problem in Section~\ref{sec:equiv}. Resynchronizations
are considered in Section~\ref{sec:resynch}. 

 \section{Preliminaries}\label{sec:preliminaries}

Given a word $w=a_1\dots a_n$, we denote by $\dom(w)=\{1,\dots,n\}$ its domain,
and by $w(i)$ its $i$-th letter, for any $i\in\dom(w)$.

\paragraph{Automata.}
To define two-way automata, and later two-way transducers, 
it is convenient to adopt the convention that, for any given input $w\in\Sigma^*$,
$w(0)=\lftmark$ and $w(|w|+1)=\rgtmark$, where 
$\lftmark,\rgtmark\notin\S$ are special markers 
used as delimiters of the input.
In this way an automaton can detect when an endpoint 
of the input has been reached and avoid moving the head 
outside.

A \emph{two-way automaton} (\emph{\NFA} for short) 
is a tuple $\Aa=(Q,\S,\D,I,F)$, where 
$\S$ is the input alphabet,
$Q=\Qleft \cupdot \Qright$ is the 
set of states, partitioned into a set $\Qleft$ of left-reading states 
and a set $\Qright$ of right-reading states, $I\subseteq \Qright$ is 
the set of initial states, $F\subseteq Q$ is the set of final states, and
$\Delta \subseteq Q\times (\Sigma\uplus\{\lftmark,\rgtmark\}) \times Q \times \{\lft,\rgt\}$ 
is the transition relation. 
The partitioning of the set of states is useful for specifying which
letter is read from each state: left-reading states read the letter 
to the left, whereas right-reading states read the letter to the right.
A transition $(q,a,q',d)\in\D$ is \emph{leftward} (resp.~\emph{rightward})
if $d=\lft$ (resp.~$d=\rgt$). 
Of course, we assume that no leftward transition is possible when reading 
the left marker $\lftmark$, and no rightward transition is possible when 
reading the right marker $\rgtmark$.
We further restrict $\D$ by asking that  $(q,a,q',\lft) \in \D$ implies 
$q' \in \Qleft$, and $(q,a,q',\rgt) \in \D$ implies $q' \in\Qright$.

To define runs of \NFA we need to first introduce the
notion of configuration.
Given a \NFA $\Aa$ and a word $w\in\Sigma^*$,  
a \emph{configuration} of $\Aa$ on $w$ is a pair $(q,i)$,
with $q\in Q$ and $i\in \{1,\ldots,|w|+1\}$. 
Such a configuration represents the fact that the automaton 
is in state $q$ and its head is {\sl between} the $(i-1)$-th 
and the $i$-th letter of $w$ (recall that we are assuming 
$w(0)=\lftmark$ and $w(|w|+1)=\rgtmark$).
The transitions that depart from a configuration 
$(q,i)$ and read $a$ are denoted $(q,i) \act{a} (q',i')$, 
and must satisfy one of the following conditions:
\begin{itemize}
\item $q \in \Qright$, $a=w(i)$, $(q,a,q',\rgt) \in\D$, and $i'=i+1$,
\item $q \in \Qright$, $a=w(i)$, $(q,a,q',\lft) \in\D$, and $i'=i$,
\item $q \in\Qleft$, $a=w(i-1)$, $(q,a,q',\rgt) \in\D$, and $i'=i$,
\item $q \in\Qleft$, $a=w(i-1)$, $(q,a,q',\lft) \in\D$, and $i'=i-1$.
\end{itemize}

\par\noindent
\begin{wrapfigure}{r}{6cm}
\vspace{-5mm}\clearpage{}\ \ \ \scalebox{0.7}{\begin{tikzpicture}[xscale=0.9]
  \draw (2,2) node [state, fill=red!25] (A) {$q_0$};
  \draw (4,2) node [state, fill=red!25] (B) {$q_1$};
  \draw (6,2) node [state, fill=red!25] (C) {$q_2$};
  \draw (6,3) node [state, fill=red!25] (D) {$q_3$};
  \draw (4,3) node [state, fill=red!25] (E) {$q_4$};
  \draw (2,3) node [state, fill=red!25] (F) {$q_5$};
  \draw (2,4) node [state, fill=red!25] (G) {$q_6$};
  \draw (4,4) node [state, fill=red!25] (H) {$q_7$};
  \draw (6,4) node [state, fill=red!25] (I) {$q_8$};
  \draw (8,4) node [state, fill=red!25] (J) {$q_9$};
   
  \draw (A) edge [arrow] node [above] {\small $a_1$ } (B) ;
  \draw (B) edge [arrow] node [above] {\small $a_2$ } (C) ;
  \draw (C) edge [arrow, bend right=90, looseness=2] node [right] {\small $a_3$ } (D) ;
  \draw (D) edge [arrow] node [above] {\small $a_2$ } (E) ;
  \draw (E) edge [arrow] node [above] {\small $a_1$ } (F) ;
  \draw (F) edge [arrow, bend left=90, looseness=2] node [left] {\small $\lftmark$ } (G) ;
  \draw (G) edge [arrow] node [above] {\small $a_1$ } (H) ;
  \draw (H) edge [arrow] node [above] {\small $a_2$ } (I) ;
  \draw (I) edge [arrow] node [above] {\small $a_3$ } (J) ;
    
  \draw (2,1) node {$1$};
  \draw (4,1) node {$2$};
  \draw (6,1) node {$3$};
  \draw (8,1) node {$4$};

  \draw (1,0) node [letter, fill=blue!25] {$\lftmark$};
  \draw (3,0) node [letter, fill=blue!25] {$a_1$};
  \draw (5,0) node [letter, fill=blue!25] {$a_2$};
  \draw (7,0) node [letter, fill=blue!25] {$a_3$};
  \draw (9,0) node [letter, fill=blue!25] {$\rgtmark$};
    
\end{tikzpicture}}\clearpage{}
\vspace{-9mm}\end{wrapfigure}
A configuration $(q,i)$ on $w$ is \emph{initial} (resp.~\emph{final})
if $q\in I$ and $i=1$ 
(resp.~$q\in F$ and $i=|w|+1$). 
A \emph{run} of $\Aa$ on $w$ is a sequence 
$\r=(q_1,i_1) \act{b_1} (q_2,i_2) \act{b_2} \cdots \act{b_m} (q_{m+1},i_{m+1})$ 
of configurations connected by transitions. 
The figure to the right
depicts an input $w=a_1 a_2 a_3$ (in blue) and a possible run on it (in red),
where $q_0,q_1,q_2,q_6,q_7,q_8\in\Qright$ and $q_3,q_4,q_5\in\Qleft$,
and $1,2,3,4$ are the positions associated with the various configurations.
A run is \emph{successful} if it starts with an initial configuration 
and ends with a final configuration. 
The \emph{language} of $\Aa$ is the set $\sem{\Aa}\subseteq\Sigma^*$ of
all words on which $\Aa$ has a successful run.

When $\Aa$ has only right-reading states (i.e.~$\Qleft=\emptyset$)
and rightward transitions, we say that $\Aa$ is a \emph{one-way automaton} 
(\emph{NFA} for short).

\paragraph{Transducers.}
Two-way transducers are defined similarly to two-way automata, by
introducing an output alphabet $\Gamma$ and associating an output 
from $\Gamma^*$ with each transition rule. 
So a \emph{two-way transducer} (\emph{\NFT} for short) 
$\Tt=(Q,\S,\G,\D,I,F)$
is basically a \NFA as
above, but with a transition relation 
$\Delta \subseteq Q\times (\Sigma\uplus\{\lftmark,\rgtmark\}) \times \Gamma^* \times Q
 \times \{\lft,\rgt\}$. 
A transition is usually denoted by $(q,i) \act{a|v} (q',i')$,
and describes a move of the transducer from configuration $(q,i)$ 
to configuration $(q',i')$
that reads the input letter $a$ and outputs the word $v$. 
The same restrictions and conventions for two-way automata apply 
to the transitions of two-way transducers, and configurations and runs 
are defined in a similar way. 

The \emph{output} associated with a successful run
$\rho=(q_1,i_1)\act{b_1\mid v_1} (q_2,i_2) \act{b_2 \mid v_2} \cdots \act{b_m \mid v_m} (q_{m+1},i_{m+1})$ 
is the word
$v_1 v_2 \cdots v_m\in\Gamma^*$. 
A two-way transducer $\Tt$ defines a relation $\sem{\Tt}\subseteq\S^*\times\Gamma^*$ 
consisting of all the pairs $(u,v)$ such that $v$ is the output of some successful 
run $\rho$ of $\Tt$ on $u$. 
Throughout the paper, a transducer is non-deterministic and two-way.

\paragraph{Origin semantics.}
In the origin semantics for transducers \cite{boj14icalp}, the output is tagged with
information about the position of the input where it was produced. If reading the $i$-th letter of the input we output $v$, 
then all letters of $v$ are tagged with $i$, and we say 
they have \emph{origin} $i$. 
We use the notation $(v,i)$ to denote that all positions in $v$ have origin $i$. 
The outputs associated with a successful run 
$\rho= (q_1,i_1) \act{b_1 \mid v_1} (q_2,i_2) \act{b_2 \mid v_2}  
       (q_3,i_3) \cdots \act{b_m \mid v_m} (q_{m+1},i_{m+1})$ 
in the origin semantics are the words of the form 
$\nu=(v_1,j_1) (v_2,j_2) \cdots (v_m,j_m)$ over $\G \times \Nat$, 
where, for all $1\le k\le m$, $j_k=i_k$ if $q_k\in \Qright$, and 
$j_k=i_k-1$ if $q_k\in \Qleft$.
Under the origin semantics, the relation defined by $\Tt$, 
denoted $\sem{\Tt}_o$, is the set of pairs $\oelement=(u,\nu)$
---called \emph{synchronized pairs}--- such that
$u\in\Sigma^*$ and $\nu\in (\G \times \Nat)^*$ 
is the output of some successful run on $u$. 
Take as example the \NFA run depicted in the previous figure, and assume 
that any transition on a letter $a \in \S$ outputs $a$, while a transition
on a marker $\lftmark$ or $\rgtmark$ outputs the empty word $\e$. 
The output associated with that run in the origin semantics is 
$(a_1,1) (a_2,2) (a_3,3) (a_2,2) (a_1,1) (a_1,1) (a_2,2)(a_3,3)$. 

Given two transducers $\Tt_1,\Tt_2$, we say they are
\emph{origin-equivalent} if $\sem{\Tt_1}_o=\sem{\Tt_2}_o$.  Note
that two transducers $\Tt_1,\Tt_2$ can be equivalent in the classical
semantics, i.e.~$\sem{\Tt_1}=\sem{\Tt_2}$, while they can have
different origin semantics, so $\sem{\Tt_1}_o\neq\sem{\Tt_2}_o$.

\paragraph{Regular outputs.}
The transducers we defined just above 
consume input letters while outputting strings of bounded length.
In order to perform some crucial constructions later ---notably,
to shortcut factors of runs with empty output--- we need to slightly 
generalize the notion of output associated with a transition, so as 
to allow producing arbitrarily long words on reading a single letter.
Formally, the transition relation of a transducer \emph{with regular outputs}
is allowed to be any subset $\D$ of 
$Q \times (\S\uplus\{\lftmark,\rgtmark\})\times \pmb{2^{\G^*}} 
   \times Q \times \{\lft,\rgt\}$
such that,
for all $q,q'\in Q$, $a\in\Sigma$, $d\in\{\lft,\rgt\}$,
there is at most one language $L\subseteq\G^*$ such that $(q,a,L,q',d)\in\D$;
moreover, this language $L$ must be non-empty and regular.
A transition $(q,i)\act{a|L} (q',i')$ 
means that the transducer can move from configuration $(q,i)$ to 
configuration $(q',i')$ while reading $a$ and outputting any word $v\in L$. 
Accordingly, the outputs that are associated with a successful run 
$\rho= (q_1,i_1) \act{b_1\mid L_1} (q_2,i_2)  \act{b_2 \mid L_2}  
       (q_3,i_3) \cdots \act{b_n \mid L_m} (q_{m+1},i_{m+1})$ 
in the origin semantics are the words of the form
$\nu=(v_1,j_1) (v_2,j_2) \cdots (v_m,j_m)$, 
where $v_k \in L_k$ and $j_k=i_k$ or $j_k=i_{k-1}$ depending on whether
$q_k\in \Qright$ or $q_k\in \Qleft$. 
We say that two runs $\r_1,\r_2$ are
\emph{origin-equivalent} if they have the same sets of associated outputs.
Clearly, the extension with regular outputs is only syntactical, and it preserves the
expressiveness of the class of transducers we consider. In the remaining of 
the paper, we will tacitly refer to above notion of transducer.

\paragraph{\pmb{\sc PSpace}-constructibility.}
As usual, we call the \emph{size} of an automaton or a transducer 
the number of its states, input symbols, transitions, 
plus, if present, the sizes of the NFA descriptions of the regular output 
languages associated with each transition rule.

In our complexity analysis, however, we will often need to work with
online presentations of automata and transducers. 
For example, we may say that an automaton or a transducer can be computed 
using a polynomial amount of working space (at thus its size would be at most 
exponential) w.r.t.~a given input. The terminology introduced below will be
extensively used throughout the paper to describe the computational complexity 
of an automaton, a transducer, or a part of it, in terms of a specific parameter.

Given a parameter $n\in\Nat$, we say that an automaton or a transducer 
has \emph{\PSPACE-constructible transitions w.r.t.~$n$} if its transition
relation can be enumerated by an algorithm that uses working space 
polynomial in $n$, 
and in addition, when the device is a transducer, every transition 
has at most polynomial size in $n$. In particular, if a transducer
has \PSPACE-constructible transitions, then the size of the NFA
representing every output language is polynomial.
Similarly, we say that an automaton is
\emph{\PSPACE-constructible w.r.t.~$n$} if all its components, 
---the alphabets, the state set, the transition relation, etc.---
are enumerable by algorithms that use space polynomial in $n$.

 \section{Equivalence of transducers with origins}\label{sec:equiv}

We focus on the equivalence problem for two-way transducers. In the classical
semantics, this problem is known to be undecidable even
if transducers are one-way~\cite{gri68} (called NGSM in the latter paper). We consider this
problem in the origin semantics. We will show that, in this setting, 
equivalence becomes decidable, and can even be solved in \PSPACE~--~so with
no more cost than equivalence of non-deterministic two-way automata:

\begin{theorem}\label{thm:main}
Containment and equivalence of two-way transducers 
under the origin semantics is \PSPACE-complete.
\end{theorem}

The proof of this result is quite technical.
As a preparation, we first show how to check origin-equivalence 
for transducers in which all transitions produce non-empty outputs,
namely, where the transition relation is of the form
$\D \subseteq 
 Q \times (\S\uplus\{\lftmark,\rgtmark\})\times \pmb{2^{\G^+}} \times Q \times \{\lft,\rgt\}$.
We call \emph{busy} any such transducer.

\subsection{Origin-equivalence of busy transducers}\label{ssec:busy}

An important feature of our definition of transducers is that, 
along any possible run, an input position is never read twice in a
row. In other words, our transducers do not have ``stay''
transitions. 
For a busy transducer, this implies that the origins of outputs of 
consecutive transitions are always different. As a consequence, runs
of two busy transducers can be only origin-equivalent if they visit the same
sequences of positions of the input and have the same possible outputs 
transition-wise. To give the intuition, we note that origin-equivalence of
busy classical transducers, with single output words associated with
transitions, can be reduced to a version of equivalence of
\NFA, where we ask that runs have the same shape. Already the last condition
does not allow to apply  the \PSPACE algorithm for
equivalence of \NFA of \cite{vardi89twoway}, 
and the naive algorithm would be of exponential time. 
Some more complications arise when we assume regular outputs, which will be
required when we shall deal with non-busy transducers.
We introduce now the key notions of transition shape and witness procedure.

Let $\Tt_1,\Tt_2$ be busy transducers over the same input alphabet,
with $\Tt_i=(Q_i,\S,\G_i,\D_i,I_i,F_i)$ for $i=1,2$. 
We say that two transitions $t_1\in\D_1$ and $t_2\in\D_2$,
with $t_i=(q_i,a_i,q'_i,L_i,d_i)$, have the \emph{same shape}
if $a_1=a_2$, 
$q_1 \in \Qlefta \iff q_2 \in \Qleftb$, and  
$q'_1 \in \Qlefta \iff q'_2  \in \Qleftb$ (and hence $d_1=d_2$).

We assume that there is a
non-deterministic procedure $\Ww$, called \emph{witness procedure},
that does the following. Given a transition
$t_1=(q_1,a_1,q'_1,L_1,d_1)$ of $\Tt_1$, 
$\Ww$ returns a set $X \subseteq \D_2$ of transitions of $\Tt_2$
satisfying the following property:
{\sl for some word $v \in L_1$}, we have 
\[
  X ~=~ 
  \big\{t_2=(p_2,a_2,q_2,L_2,d_2) \in \D_2 \::\: 
        v \in L_2, \text{ and $t_2$ has {\sl same shape} as $t_1$} \big\}.
\]
Note that $\Ww$ is non-deterministic: it can return several sets
based on the choice of $v$. 
If $t_1 \in \D_1$ and $X$ is a set that could be returned by $\Ww$ on $t_1$, we write $X \in \Ww(t_1)$. However, if $\Tt_1$ and $\Tt_2$ were classical transducers, 
specifying a single output word for each transition, 
then $\Ww$ could return only one set on $t_1 \in \D_1$, that is, the set of transitions of $\Tt_2$ with the same shape and the same output as $t_1$.

The intuition behind the procedure $\Ww$ is the following. 
Consider a successful run $\r_1$ of $\Tt_1$ on $u$. 
Since $\Tt_1,\Tt_2$ are both busy,  $\sem{\Tt_1}_o \subseteq \sem{\Tt_2}_o$ 
necessarily means that for all possible outputs produced by the transitions 
of $\r_1$, there is some run $\r_2$ of $\Tt_2$ on $u$ that has the same shape 
as $\r_1$ and the same outputs, transition-wise. 
Procedure $\Ww$ will precisely provide, for each transition $t_1$ of $\Tt_1$ 
with output language $L \subseteq \G^+$, and for each choice of $v \in L$,
the set of all transitions of $\Tt_2$ with the same shape as $t$ and that could produce the same output $v$.

We introduce a last piece of terminology.
Given a run $\r_1=t_1\dots t_m$ of $\Tt_1$ of length $m$
and a sequence $\xi=X_1,\ldots,X_m$ of subsets of $\D_2$
(\emph{witness sequence}), we write $\xi \in \Ww(\r_1)$ 
whenever $X_i \in \Ww(t_i)$ for all $1\le i\le m$.
We say that a run $\r_2=t'_1 \dots t'_m$ of $\Tt_2$ is 
\emph{$\xi$-compatible} if $t'_i\in X_i$ for all $1\le i\le m$.
The following result is crucial:
    
\begin{proposition}\label{prop:from-containment-to-witness}
Let $\Tt_1,\Tt_2$ be two busy transducers over the same input alphabet,
and $\Ww$ a witness procedure. 
Then $\sem{\Tt_1}_o \subseteq \sem{\Tt_2}_o$ 
if and only if for every successful run $\r_1$ of $\Tt_1$, and for every witness sequence $\xi \in \Ww(\r_1)$,
there is a successful run $\r_2$ of $\Tt_2$ which is $\xi$-compatible.
\end{proposition}

\begin{proof}
Let $\Tt_i=(Q_i,\S,\G_i,\D_i,I_i,F_i)$ for $i=1,2$. 
We first assume that $\sem{\Tt_1}_o \subseteq \sem{\Tt_2}_o$. 
Consider a successful run $\r_1=t_1 \dots t_m$ of $\Tt_1$ on input $u$,
with each transition $t_k$ of the form $(q_k,i_k) \act{b_k\mid L_k} (q_{k+1},i_{k+1})$.
Choose any witness sequence $\xi = X_1,\dots,X_m$ in $\Ww(\r_1)$.
Recall that each $X_k$ corresponds to a choice of an output $v_k \in L_k$. 
So $\nu=(v_1,j_1) \cdots (v_m,j_m)$ is an output produced by $\r_1$,
where each $j_k$ is either $i_k$ or $i_k-1$ depending on whether $q_k$ is
right-reading or left-reading.
Since $\sem{\Tt_1}_o \subseteq \sem{\Tt_2}_o$, 
there must be a successful run $\r_2=t'_1 \dots t'_m$ of $\Tt_2$ 
on the same input $u$
that enables the same $\nu$ as output.
In other words, for this $\r_2=t'_1,\ldots,t'_m$, we have $t'_k \in X_k$, 
since the output language of $t'_k$ contains $v_k$. This shows that
$\r_2$ is $\xi$-compatible.

For the converse implication, let $\r_1=t_1 \dots t_m$ be a successful run 
of $\Tt_1$ on $u$, with $L_k$ output language of $t_k$, for all $1\le k\le m$,
and consider a possible output $\nu=(v_1,j_1) \cdots (v_m,j_m)$ produced by $\r_1$.
We want to show that $\nu$ can also be produced by a successful run of $\Tt_2$ on
the same input $u$. 
According to the description of the witness procedure $\Ww$, 
for each $k$ there is a set $X_k\in\Ww(t_k)$ containing precisely 
the transitions $t'$ of $\Tt_2$ that can output $v_k$ and that have
the same shape as $t_k$. 
Let $\xi=X_1,\ldots,X_m$. By the hypothesis of the claim, there is a 
successful run $\r_2=t'_1 \dots t'_m$ of $\Tt_2$ that is $\xi$-compatible.
This means that the transitions in $\r_2$ have the same shapes as those in 
$\r_2$ ---in particular, they read the same input letters--- and can produce 
the same outputs $v_1,\dots,v_m$, tagged with the same origins $j_1,\dots,j_m$.
Thus, $\r_2$ is over the same input $u$ and enables the same
output $\nu=(v_1,j_1) \cdots (v_m,j_m)$ as $\r_1$.
\end{proof}

Next, we reduce the problem $\sem{\Tt_1}_o \subseteq \sem{\Tt_2}_o$ 
to the emptiness problem of a one-way automaton (NFA) $\Bb$.
In this reduction, the NFA $\Bb$ can be exponentially larger than $\Tt_1,\Tt_2$, 
but is \PSPACE-constructible under suitable assumptions on $\Tt_1,\Tt_2$, and $\Ww$.

\begin{lemma}\label{lemma:from-witness-to-nfa}
Given two busy transducers $\Tt_1,\Tt_2$ with input alphabet $\S$
and a witness procedure $\Ww$, one can construct an NFA $\Bb$ that 
accepts precisely the words $u\in\S^*$ 
for which there exist a successful run $\r_1$ of $\Tt_1$ on $u$
and a witness sequence $\xi\in\Ww(\r_1)$ such that {\sl no}
$\xi$-compatible run $\r_2$ of $\Tt_2$ is successful.

Moreover, if $\Tt_1,\Tt_2$ have a total number $n$ of states
and \PSPACE-constructible transitions w.r.t.~$n$, 
and $\Ww$ uses space polynomial in $n$, 
then $\Bb$ is \PSPACE-constructible w.r.t.~$n$.
\end{lemma}

Before proving the lemma, let us state an immediate consequence
of it and of the previous Proposition \ref{prop:from-containment-to-witness}:

\begin{corollary}\label{cor:busy-containment}
Given two busy transducers $\Tt_1,\Tt_2$ with a total number $n$ of states, 
and given a witness procedure that uses space polynomial in $n$,
the problem of deciding  $\sem{\Tt_1}_o \subseteq \sem{\Tt_2}_o$ 
is in \PSPACE w.r.t.~$n$.
\end{corollary}

\begin{proof}
By Proposition \ref{prop:from-containment-to-witness}, 
$\sem{\Tt_1}_o \subseteq \sem{\Tt_2}_o$ amounts to 
check emptiness of the NFA $\Bb$ from Lemma \ref{lemma:from-witness-to-nfa}. 
The latter problem can be decided in \PSPACE w.r.t.~$n$, 
by using the constructibility of $\Bb$ to enumerate all 
transitions departing from any given state.
\end{proof}

We finally turn to the proof of the lemma:

\begin{proof}[Proof of Lemma~\ref{lemma:from-witness-to-nfa}]
Let $\Tt_i=(Q_i,\S,\G_i,\D_i,I_i,F_i)$ for $i=1,2$.
The goal is as expected: on input $u$, the NFA $\Bb$
needs to verify the existence of a successful run
$\r_1=t_1,\ldots,t_m$ of $\Aa_1$ on $u$, and of a witness sequence
$\xi \in \Ww(\r_1)$, such that there is no successful,
$\xi$-compatible run $\r_2=t'_1,\ldots,t'_m$ of $\Aa_2$
(the fact that $\r_2$ is over the same input $u$ as $\r_1$
follows, as usual, from $\xi$-compatibility).
The goal is achieved by processing the input $u$ from left to right,
while guessing pieces of the run $\r_1$ of $\Tt_1$ that are `induced' 
by prefixes of $u$. At the same time, $\Bb$ uses the procedure
$\Ww$ to `guess' the witness sequence $\xi$. 
The sequence $\xi$ is used to track induced pieces of runs of 
$\Tt_2$ with the same shape as in $\r_1$, such that once we
choose an output word for every transition of $\Tt_1$, we are
guaranteed to follow all pieces of runs of $\Tt_2$ that can 
produce the same output words.

Recall that successful runs  start 
at the leftmost position and end at the rightmost position. 
We say that a run is \emph{induced} by the first $i$ letters 
of the input $u$ if it visits only the positions $\{1,\ldots,i\}$,
possibly ending in position $i+1$
(in particular, if $i\le |u|$, then such a run will never read 
the right marker $\rgtmark$).
We further restrict our attention to induced runs that end 
at the rightmost position, i.e.~$i+1$, and that start at any 
of the two extremities, i.e.~$1$ or $i+1$.
Based on the starting position, we distinguish between 
\emph{left-to-right} runs and \emph{right-to-right} runs. 
For example, if we refer to the example of run given in Section \ref{sec:preliminaries}
and consider the first two letters of the input, we would 
have one induced left-to-right run 
$(q_0,1) \act{a_1} (q_1,2) \act{a_2} (q_2,3)$,
and one induced right-to-right run 
$(q_3,3) \act{a_2} (q_4,2) \act{a_1} (q_5,1) \act{\lftmark} 
 (q_6,1) \act{a_1} (q_7,2) \act{a_2} (q_8,3)$.

The NFA $\Bb$ will guess runs of $\Tt_1$ induced by 
longer and longer prefixes of the input (more precisely, for each
prefix, there will be exactly one left-to-right run starting at 
an initial state and any number of right-to-right runs).
At the end, $\Bb$ expects to see only one left-to-right induced 
run of $\Tt_1$, that must be successful.
At the same time, $\Bb$ will also maintain the set of all runs of 
$\Tt_2$ that are induced by the consumed prefix and that have been
constructed using transitions provided by $\Ww$. 
Concerning these latter runs, $\Bb$ will check that none 
of the left-to-right runs of $\Tt_2$ are successful. 
In fact, as we will see below, all the guessed runs are 
abstracted by states or pairs of states, so as to maintain 
at each step only a bounded, polynomial-size information.
This will require some care, since, for instance, the same pair 
of states can represent many right-to-right runs of $\Tt_1$,
with different shapes. 
The key observation is that it is sufficient to associate at 
most one left-to-right run with each state of $\Tt_1$, and 
at most one right-to-right run with each pair of states of
$\Tt_1$. This is because if, say, $\r_1$ has two right-to-right runs
on some prefix, that start and end, respectively, in the same states,
then we can replace one run by the other, and do the same for
$\r_2$. The property that no $\xi$-compatible run of $\Tt_2$ is
ultimately successful, will be preserved.
  
The states of the NFA $\Bb$ are tuples consisting of four different objects
(note that the size of a state is polynomial in the total number $n$ of states of 
$\Tt_1,\Tt_2$):
\begin{itemize}
  \item a right-reading state $\hat q$ of $\Tt_1$,
  \item a set $P$ of pairs of left-reading and right-reading states of $\Tt_1$,
  \item a set $S$ of right-reading states of $\Tt_2$, 
  \item a relation $R$ between pairs of states of $\Tt_1$
        and pairs of states of $\Tt_2$ such that
        for all $\big((q_1,q'_1),(q_2,q'_2)\big)\in R$, 
        $q_1$ and $q_2$ are left-reading, and 
        $q'_1$ and $q'_2$ are right-reading.
\end{itemize}
The intended meaning of the above objects is explained by the following invariant.
After reading a prefix $u[1,i]$ of the input, the state
$(\hat q,P,S,R)$ reached by $\Bb$ must satisfy the following properties:
\begin{enumerate}
  \item there is a left-to-right run $\r_{\hat q}$ of $\Tt_1$ 
        induced by $u[1,i]$ and ending in $\hat q$, and a
        witness sequence $\xi_{\hat q}$ in $\Ww(\r_{\hat q})$,
  \item for each pair $(q,q')\in P$, 
        there is a right-to-right run $\r_{q,q'}$ of $\Tt_1$ that is
        induced by $u[1,i]$, starts in $q$, ends in $q'$, and a
        witness sequence $\xi_{q,q'}$ in $\Ww(\r_{q,q'})$,
  \item if there is a left-to-right run $\r'$ of $\Tt_2$ that is induced by
        $u[1,i]$ and is $\xi_{\hat q}$-compatible (cf.~first item), 
        then $S$ contains the last state of $\r'$,
  \item for each pair $(q,q')\in P$, 
        if there is a right-to-right run $\r'$ of $\Tt_2$ 
        induced by $u[1,i]$, $\xi_{q,q'}$-compatible
        (cf.~second item), starting in $r$ and ending in $r'$,
        then $((q,q'),(r,r'))\in R$.
\end{enumerate}
The initial states of $\Bb$ are the tuples of the form
$(\hat q,P,S,R)$, with $\hat q\in I_1$, $P=\emptyset$, and
$S=I_2$. Similarly, the final states are the tuples $(\hat q,P,S,R)$ 
such that $\hat q\in F_1$, $P=\emptyset$, and $S\cap F_2 = \emptyset$.
Assuming that Properties 1.--4.~are satisfied, this will imply
that $\Bb$ accepts some input $u$ iff there exist a successful
run $\r_1$ of $\Tt_1$ on $u$ and a witness sequence $\xi \in \Ww(\r_1)$, 
but no successful, $\xi$-compatible run of $\Tt_2$ on $u$.

We now give the transitions of $\Bb$ 
that preserve Properties 1.--4.
These are of the form
\[
  (\hat q,P,S,R) ~\act{a}~ (\hat q',P',S',R')
\]
and must satisfy a certain number of constraints between
the various components of the source and target states.
We first focus on the constraints between the first two 
components, i.e.~$\hat q,P$ and $\hat q',P'$, which 
guarantee Properties 1.--2. 
For this, we basically apply a variant of the classical 
crossing-sequence construction \cite{she59} for 
simulating a \NFA by an NFA. In the following we omit output languages
in transitions, since they are determined by the source/target state,
the letter and the direction.

\begin{itemize}
\item
A first condition requires that $\hat q$ is connected to $\hat q'$ 
by a sequence of $k$ leftward transitions on $a$ interleaved by $k$ 
right-to-right induced runs, and followed by a single rightward 
transition on $a$, as follows 
(see also the left hand-side of Figure \ref{fig:shape-check}):
\[
  (\hat q,a,q_1,\lft) ~~ \r_{q_1,q_2} ~~ 
  (q_2,a,q_3,\lft) ~~ \r_{q_3,q_4} ~~ \dots ~~
  \r_{q_{2k-1},q_{2k}} ~~
  (q_{2k},a,\hat q',\rgt)
\]
where $(q_1,q_2),(q_3,q_4),\dots,(q_{2k-1},q_{2k}) \in P$.
\par
For brevity, we denote this property by 
$\hat q \leadsto q_1 \leadsto q_2 \dots \leadsto q_{2k} \leadsto \hat q'$.

\begin{figure}[!t]
\centering
\clearpage{}\vspace{-2mm}
\scalebox{0.6}{
\begin{tikzpicture}
\begin{scope}[xshift=0cm,scale=0.75]
  \draw (0,0) node (A) {};
  \draw (4,0) node (B) {$\hat q$};
  \draw (4,1) node (C) {$q_1$};
  \draw (4,2) node (D) {$q_2$};
  \draw (4,3) node (E) {\LARGE $^{\vdots}$};
  \draw (4,4) node (F) {$q_{2k-1}$};
  \draw (4,5) node (G) {$q_{2k}$};
  \draw (6,5) node (H) {$\hat q'$};
   
  \draw (A) edge [arrow, gray, dotted] node [above left] {$\rho_{\hat q}$} (B);
  \draw (B) edge [arrow, bend right=90, looseness=2] node [right] {$a$} (C);
  \draw (C) -- (2.25,1);
  \draw (2.5,1) edge [bend left=90, looseness=2] (2.5,2);
  \draw [arrow] (2.5,2) -- node [above left] {$\rho_{q_1,q_2}$} (D);
  \draw (D) edge [arrow, bend right=90, looseness=2] node [right] {$a$} (E);
  \draw (F) -- (2,4);
  \draw (2,4) edge [bend left=90, looseness=2] (2,5);
  \draw [arrow] (2,5) -- node [above left] {$\rho_{q_{2k-1},q_{2k}}$} (G);
  \draw (G) edge [arrow] node [above] {$a$} (H);
  
  \draw [thin, dashed, gray] (4,-1) -- (B);
  \draw [thin, dashed, gray] (B) -- (C);
  \draw [thin, dashed, gray] (C) -- (D);
  \draw [thin, dashed, gray] (D) -- (E);
  \draw [thin, dashed, gray] (E) -- (F);
  \draw [thin, dashed, gray] (F) -- (G);
  \draw [thin, dashed, gray] (G) -- (4,6);
  \draw [thin, dashed, gray] (6,-1) -- (H);
  \draw [thin, dashed, gray] (H) -- (6,6);
\end{scope}
\begin{scope}[xshift=10cm,yshift=-0.5cm,scale=0.75]
  \draw (6,1) node (B) {$q$};
  \draw (4,1) node (C) {$q_1$};
  \draw (4,2) node (D) {$q_2$};
  \draw (4,3) node (E) {\LARGE $^{\vdots}$};
  \draw (4,4) node (F) {$q_{2k-1}$};
  \draw (4,5) node (G) {$q_{2k}$};
  \draw (6,5) node (H) {$q'$};
   
  \draw (B) edge [arrow] node [above] {$a$} (C);
  \draw (C) -- (2.5,1);
  \draw (2.5,1) edge [bend left=90, looseness=2] (2.5,2);
  \draw [arrow] (2.5,2) -- node [above left] {$\rho_{q_1,q_2}$} (D);
  \draw (D) edge [arrow, bend right=90, looseness=2] node [right] {$a$} (E);
  \draw (F) -- (2,4);
  \draw (2,4) edge [bend left=90, looseness=2] (2,5);
  \draw [arrow] (2,5) -- node [above left] {$\rho_{q_{2k-1},q_{2k}}$} (G);
  \draw (G) edge [arrow] node [above] {$a$} (H);
  
  \draw [thin, dashed, gray] (4,0) -- (C);
  \draw [thin, dashed, gray] (C) -- (D);
  \draw [thin, dashed, gray] (D) -- (E);
  \draw [thin, dashed, gray] (E) -- (F);
  \draw [thin, dashed, gray] (F) -- (G);
  \draw [thin, dashed, gray] (G) -- (4,6);
  \draw [thin, dashed, gray] (6,0) -- (B);
  \draw [thin, dashed, gray] (B) -- (H);
  \draw [thin, dashed, gray] (H) -- (6,6);
\end{scope}
\end{tikzpicture}
}
\vspace{-6mm}
\clearpage{}
\caption{Constraints between $\hat q$ and $\hat q'$, 
         and between $P$ and $P'$.}\label{fig:shape-check}
\end{figure}
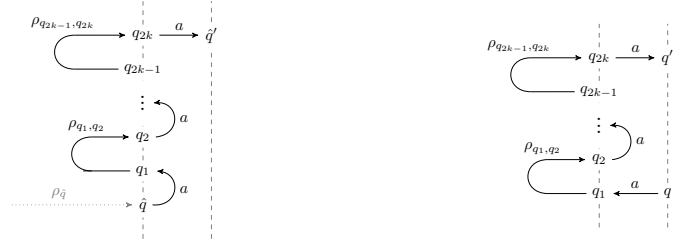

\item 
In a similar way, we require that, for every pair $(q,q')\in P$,
$q$ is connected to $q'$ by a leftward transition on $a$,
then a sequence of $k>0$ right-to-right induced runs interleaved 
by $n-1$ leftward transitions on $a$, followed by
a rightward transition on $a$, as follows
(see the right hand-side of Figure \ref{fig:shape-check}):
\[
  (q,a,q_1,\lft) ~~
  \r_{q_1,q_2} ~~ (q_2,a,q_3,\lft) ~~ 
  \r_{q_3,q_4} ~~ \dots ~~
  \r_{q_{2k-1},q_{2k}} ~~
  (q_{2k},a,q',\rgt)
\]
where $(q_1,q_2),(q_3,q_4),\dots,(q_{2k-1},q_{2k}) \in P$.
\par
As before, we write for short 
$q \leadsto q_1 \leadsto q_2 \dots \leadsto q_{2k} \leadsto q'$
(to distinguish the notations for the former and the latter property, 
it suffices to check whether the first state is right-reading or left-reading).
\end{itemize}

As concerns the constraints between $S,R$ and $S',R'$, we 
first lift the previous properties and notations to pairs 
of runs of $\Tt_1$ and $\Tt_2$, thus writing, for instance,
$\bin{\hat q}{s} \binleadsto \bin{q_1}{s_1} 
                 \binleadsto \bin{q_2}{s_2} \dots 
                 \binleadsto \bin{q_{2k}}{s_{2k}} 
                 \binleadsto \bin{\hat q'}{s'}$.
On top of this, we restrict the runs of $\Tt_2$ to be $\xi$-compatible
with the  runs of $\Tt_1$, for the corresponding witness sequence
$\xi$. Formally, we require the following:
\begin{itemize}
\item Assume that  $s \in S$ and that
$R$ contains the pairs $\big((q_{2i-1},q_{2i}),(s_{2i-1},s_{2i})\big)$,
for all $1\le i\le k$.
Let $t=(\hat q,a,q_1,\lft)$, $t'=(q_{2k},a,\hat q',\rgt)$, and
$t_i=(q_{2i-1},a,q_{2i},\lft)$, for all $2\le i\le k$, be the transitions
of $\Tt_1$ that are used to connect $\hat q$ to $q_1$, $q_{2k}$ to $\hat q'$, etc.

Using $\Ww$, choose some witness sets 
$X \in \Ww(t)$, $X' \in \Ww(t')$, and $X_i \in \Ww(t_i)$, for $2\le i\le k$. 
Then $s'\in S'$ if
\[
  \bin{\hat q}{s} \binleadsto \bin{q_1}{s_1} 
                  \binleadsto \bin{q_2}{s_2} \dots 
                  \binleadsto \bin{q_{2k}}{s_{2k}} 
                  \binleadsto \bin{\hat q'}{s'}
\]
for some $(s,a,s_1,\lft) \in X$, $(s_{2k},a,s',\rgt) \in X'$,
$(q_{2i-1},a,q_{2i},\lft) \in X_i$ ($2\le i\le k$).
\item Assume that $(q,q')\in P'$ 
and $R$ contains the pairs
$\big((q_{2i-1},q_{2i}),(s_{2i-1},s_{2i})\big)$, 
for all $1\le i\le k$.
As before, let $t,t',t_i$ 
be the transitions of $\Tt_1$ that connect $q$ to $q_1$, $q_{2k}$ to $q'$, 
$q_{2i}$ to $q_{2i+1}$, for all $1\le i\le k$.

Using $\Ww$, choose some witness sets 
$X \in \Ww(t))$, $X' \in \Ww(t')$, and $X_i \in \Ww(t_i)$, for $1\le i<k$. 
Then $\big((q,q'),(s,s')\big)\in R'$ if 
\[
  \bin{q}{s} \binleadsto \bin{q_1}{s_1} 
             \binleadsto \bin{q_2}{s_2} \dots 
             \binleadsto \bin{q_{2k}}{s_{2k}} 
             \binleadsto \bin{q'}{s'}
\]
for some $(s,a,s_1,\lft) \in X$, $(s_{2k},a,s',\rgt) \in X'$,
$(s_{2k},a,s_{2k+1},\lft) \in X_i$ ($1\le i<k$).
\end{itemize}
It is routine to check that the above constraints guarantee
all the Properties 1.--4.~stated above.
To conclude, we observe that the NFA $\Bb$ can be constructed by
an algorithm that uses the subprocedures for enumerating the transitions 
of $\Tt_1,\Tt_2$ and the non-deterministic \PSPACE procedure $\Ww$, 
plus additional polynomial space for storing (temporarily) the sets $X,X',X_i$.
\end{proof}

\subsection{Origin-equivalence of arbitrary transducers}\label{sec:arbitrary}

We now consider transducers that are not necessarily busy.
To show that origin-equivalence remains decidable in \PSPACE, we  
will modify the transducers so as to make them busy, 
and reduce in this way the origin equivalence problem 
to the case treated in Section \ref{ssec:busy}.

A naive idea would be to modify the transitions that output 
the empty word $\e$ and make them output a special letter $\#$.
This however would not give a correct reduction towards 
origin-equivalence with busy transducers. 
Indeed, a transducer may produce non-empty outputs, 
say $v_1,v_2,\dots$, with transitions that occur at the same 
position, say $i$, and that are interleaved by runs 
traversing other positions of the input but producing only $\e$.
In that case, we would still need to compare where the words 
$v_1,v_2,\dots$ were produced, and see that they may form a 
contiguous part of the output with origin $i$.
The above idea is however useful if we first \emph{normalize} 
our transducers in such a way that maximal subruns generating 
empty outputs are unidirectional. Paired with the fact that
the same input position is never visited twice on two consecutive
transitions, this will give the following characterization:
two arbitrary transducers are origin-equivalent if and only if 
their normalized versions, with empty outputs replaced by $\#$,  
are also origin-equivalent.

\smallskip
For simplicity, we fix a single transducer $\Tt=(Q,\S,\G,\D,I,F)$,
which could be thought of as any of the two transducers
$\Tt_1,\Tt_2$ that are tested for origin-equivalence.
To normalize $\Tt$ we consider 
runs that start and end in the same position, 
and that produce empty output. 
Such runs are called lazy U-turns and are formally 
defined below. We will then abstract lazy U-turns 
by pairs of states, called U-pairs for short.
  
\begin{definition}\label{def:Uturn}
Given an input word $u$,
a \emph{left} (resp.~\emph{right}) \emph{lazy U-turn at position $i$ of $u$} 
is any run of $\Tt$ on $u$ of the form
$(q_1,i_1) \act{b_1\mid v_1} (q_2,i_2) \act{b_2\mid v_2} \cdots \act{b_m\mid v_m} 
 (q_{m+1},i_{m+1})$,
with 
$i_1=i_{m+1}=i$, $i_k<i$ (resp.~$i_k>i$) for all $2\le k\le m$,
and $v_k=\e$ for all $1\le k\le m$. 
\end{definition}

\noindent
For brevity, we shall often refer to a \emph{left/right lazy U-turn} 
without specifying the position $i$ and the word $u$, assuming that
these are clear from the context.

The pair $(q_1,q_{m+1})$ of states at the extremities of a 
left/right lazy U-turn is called a \emph{left/right U-pair}
(\emph{at position $i$ of $u$}).
We denote by $U_i^\lefttorightarrow$ (resp.~$U_i^\righttoleftarrow$)
the set of all left (resp.~right) U-pairs at position $i$ 
(again, the input $u$ is omitted from the notation as it 
usually understood from the context). 

Note that we have
$U_i^\lefttorightarrow \subseteq \Qleft\times\Qright$ and
$U_i^\righttoleftarrow \subseteq \Qright\times\Qleft$.
Accordingly, we define the word 
$u^\lefttorightarrow$ over $2^{\Qleft\times\Qright}$ 
that has the same length as $u$ and labels every position 
$i$ with the set $U_i^\lefttorightarrow$ of left U-pairs.
This $u^\lefttorightarrow$ is seen as an annotation of the 
original input $u$, and can be computed from $\Tt=(Q,\S,\G,\D,I,F)$ 
and $u$ using the following recursive rule:

\vspace{4mm}
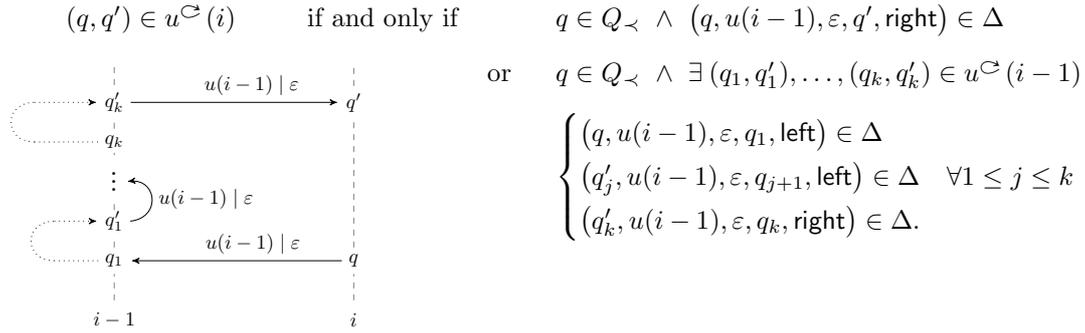
\begin{minipage}{5mm}
\vspace{6mm}\scalebox{0.75}{\hspace{-6mm}\begin{tikzpicture}[scale=0.7]
  \draw (10,1) node (B) {$q$};
  \draw (4,1) node (C) {$q_1$};
  \draw (4,2) node (D) {$q'_1$};
  \draw (4,3) node (E) {\LARGE $^{\vdots}$};
  \draw (4,4) node (F) {$q_k$};
  \draw (4,5) node (G) {$q'_k$};
  \draw (10,5) node (H) {$q'$};

  \draw (10,-.5) node (X) {$i$};
  \draw (4,-.5) node (X') {$i-1$};
   
  \draw (B) edge [arrow] node [above] {$\ \ \ \ \ u(i-1)\mid\e$} (C);
  \draw [dotted] (C) -- (2.5,1);
  \draw (2.5,1) edge [dotted, bend left=90, looseness=2] (2.5,2);
  \draw [dotted, arrow] (2.5,2) -- (D);
  \draw (D) edge [arrow, bend right=90, looseness=2] node [right] {$u(i-1)\mid\e$} (E);
  \draw [dotted] (F) -- (2,4);
  \draw (2,4) edge [dotted, bend left=90, looseness=2] (2,5);
  \draw [dotted, arrow] (2,5) -- (G);
  \draw (G) edge [arrow] node [above] {$\ \ \ \ \ u(i-1)\mid\e$} (H);
  
  \draw [thin, dashed, gray] (4,0) -- (C);
  \draw [thin, dashed, gray] (C) -- (D);
  \draw [thin, dashed, gray] (D) -- (E);
  \draw [thin, dashed, gray] (E) -- (F);
  \draw [thin, dashed, gray] (F) -- (G);
  \draw [thin, dashed, gray] (G) -- (4,6);
  \draw [thin, dashed, gray] (10,0) -- (B);
  \draw [thin, dashed, gray] (B) -- (H);
  \draw [thin, dashed, gray] (H) -- (10,6);
\end{tikzpicture}}
\end{minipage}~
\begin{minipage}{\linewidth-5mm}
\[
  \begin{array}{rl}
    \hspace{-12mm}
    (q,q')\in u^\lefttorightarrow(i)
    ~~\qquad\text{if and only if}\qquad~~
         & q\in\Qleft ~\wedge~ \big(q,u(i-1),\e,q',\rgt\big)\in\D \\[2ex]
    \text{or}~~ & q\in\Qleft ~\wedge~ \exists \: 
           (q_1,q'_1),\dots,(q_k,q'_k) \in u^\lefttorightarrow(i-1) \\[2ex]
         & \begin{cases}
             \big(q,u(i-1),\e,q_1,\lft\big) \in \D \\[0.5ex]
             \big(q'_j,u(i-1),\e,q_{j+1},\lft\big) \in \D 
             & \forall 1\le j\le k \\[0.5ex]
             \big(q'_k,u(i-1),\e,q_k,\rgt\big) \in \D.
           \end{cases}  
  \end{array}
\]
\vspace{12mm}
\end{minipage}

\par\noindent
The annotation $u^\righttoleftarrow$ 
with right $U$-pairs satisfies a symmetric recursive rule.

Of course, the annotations $u^\lefttorightarrow,u^\righttoleftarrow$
are over alphabets of exponential size. This does not raise
particular problems concerning the complexity, since we aim at
deciding origin-equivalence in \PSPACE, and towards this goal 
we could work with automata and transducers that have 
\PSPACE-constructible transitions.
In particular, we can use the recursive rules for $u^\lefttorightarrow$
and $u^\righttoleftarrow$ to get the following straightforward lemma (proof omitted):

\begin{lemma}\label{lemma:annotations}
Given a transducer $\Tt$, one can compute an NFA $\Uu$ that has the
same number of states as $\Tt$ and such that
$\sem{\Uu} = 
  \{ u\otimes u^\lefttorightarrow\otimes u^\righttoleftarrow
     \::\: u\in\Sigma^* \}$,
where $\otimes$ denotes the convolution of words of the same length.
The NFA $\Uu$ is \PSPACE-constructible w.r.t.~the size of $\Tt$.
\end{lemma}

To normalize the transducer $\Tt$ it is convenient to assume that
the sets of left and right U-pairs can be read directly from the input
(we refer to this as \emph{annotated input}).
For this, we introduce the transducer $\Tt_U$, that is obtained from 
$\Tt$ by extending its input alphabet from $\Sigma$ to 
$\Sigma\times 2^{\Qleft\times\Qright}\times 2^{\Qright\times\Qleft}$,
and by modifying the transitions in the obvious way, that is,
from $(q,a,L,q',d)$ to $\big(q,(a,U^\lefttorightarrow,U^\righttoleftarrow),L,q',d\big)$
for any $U^\lefttorightarrow\subseteq\Qleft\times\Qright$ and
$U^\righttoleftarrow\subseteq\Qright\times\Qleft$.
Note that $\Tt_U$ does not check that the input is correctly 
annotated, i.e.~of the form $u\otimes u^\lefttorightarrow\otimes u^\righttoleftarrow$  (this is done by the NFA $\Uu$). 
Further note that $\Tt_U$ has exponential size w.r.t.~$\Tt$,
but its state space remains the same as $\Tt$, and its 
transitions can be enumerated in \PSPACE.

\smallskip
We are now ready to describe the normalization of $\Tt_U$, which 
produces an origin-equivalent transducer $\norm(\Tt_U)$ with no lazy U-turns.
The transducer $\norm(\Tt_U)$ is obtained in two steps.
First, using the information provided by the annotation of the input, 
we shortcut all runs of $\Tt_U$ that consist of multiple transitions 
outputting at the same position and
\begin{wrapfigure}{r}{4.7cm}
\vspace{-8mm}\scalebox{0.6}{
\begin{tikzpicture}[xscale=0.9,yscale=0.875]
  \draw (4,1) node (A) {$q$};
  \draw (8,1) node (B) {$q'_1$};
  \draw (8,2) node (C) {$q_2$};
  \draw (8,3) node (D) {\LARGE $^{\vdots}$};
  \draw (8,4) node (E) {$q_{k-1}$};
  \draw (4,4) node (F) {$q'_{k-1}$};
  \draw (4,5) node (G) {$q_k$};
  \draw (4,6) node (H) {$q'$};
   
  \draw [red] (A) edge [arrow, line width=1mm, out=35, in=30, looseness=2.4] 
        node [above=13mm] {$a\mid L\supseteq L_1\cdots L_k\mspace{100mu}\ $} (H);

  \draw [white] (A) edge [arrow, line width=1mm] (B);
  \draw [white] (C) edge [arrow, line width=1mm, bend left=90, looseness=2] (D);
  \draw [white] (E) edge [arrow, line width=1mm] (F);
  \draw [white] (G) edge [arrow, line width=1mm, bend right=90, looseness=2] (H);
  
  \draw (A) edge [arrow, very thick] node [above] {$a\mid L_1$} (B);
  \draw [dotted] (B) -- (9.5,1);
  \draw (9.5,1) edge [dotted, bend right=90, looseness=2] (9.5,2);
  \draw [dotted, arrow] (9.5,2) -- (C);
  \draw (C) edge [arrow, very thick, bend left=90, looseness=2] node [left] {$a\mid L_2$} (D);
  \draw (E) edge [arrow, very thick] node [above] {$a\mid L_{k-1}$} (F);
  \draw [dotted] (F) -- (2.25,4);
  \draw (2.25,4) edge [dotted, bend left=90, looseness=2] (2.5,5);
  \draw [dotted,arrow] (2.25,5) -- (G);
  \draw (G) edge [arrow, very thick, bend right=90, looseness=2] node [right] {$a\mid L_k$} (H);
  
  \draw [thin, dashed, gray] (4,0) -- (A);
  \draw [thin, dashed, gray] (A) -- (F);
  \draw [thin, dashed, gray] (F) -- (G);
  \draw [thin, dashed, gray] (G) -- (H);
  \draw [thin, dashed, gray] (H) -- (4,7);
  \draw [thin, dashed, gray] (8,0) -- (B);
  \draw [thin, dashed, gray] (B) -- (C);
  \draw [thin, dashed, gray] (C) -- (D);
  \draw [thin, dashed, gray] (D) -- (E);
  \draw [thin, dashed, gray] (E) -- (8,7);
\end{tikzpicture}
}
 \vspace{-4mm}\end{wrapfigure}
interleaved by lazy U-turns.
The resulting transducer is denoted $\shortcut(\Tt_U)$.
After this 
step, we will eliminate the lazy $U$-turns, 
thus obtaining $\norm(\Tt_U)$. 
Formally, $\shortcut(\Tt_U)$ has 
for transitions 
the tuples of the form $\big(q,(a,U^{\lefttorightarrow},U^{\righttoleftarrow}),L,q',d\big)$,
where $L$ is the smallest language 
that contains every language of the form $L_1 \cdot L_2 \cdots L_k$ 
for which there 
are $q_1,q'_1,\dots,q_k,q'_k$ and $d_1,\dots,d_k$,
with 
$q = q_1$, $q'_k = q'$ (and hence $d_k=d$),
$(q_i,a,L_i,q'_i,d_i) \in \Delta$, 
and $(q'_i,q_{i+1}) \in U^\lefttorightarrow \cup\, U^\righttoleftarrow$
for all $i$
(see the figure to the 
right). 
Note that there is no transition $\big(q,(a,U^{\lefttorightarrow},U^{\righttoleftarrow}),L,q',d\big)$
when there are no languages $L_1,L_2,\dots,L_k$ as above.\linebreak 
The output languages associated with the transitions of $\shortcut(\Tt_U)$
can be constructed using a classical saturation mechanism, which is omitted 
here, they are regular, and their NFA representations are polynomial-sized 
w.r.t.~the size of the NFA representations of the output languages of $\Tt_U$. 
This implies that $\shortcut(\Tt_U)$ has \PSPACE-constructible transitions 
w.r.t.~the number of its states, exactly like $\Tt_U$.

Recall that two runs are 
\emph{origin-equivalent} if they produce the same synchronized pairs.
The next lemma shows that, on correctly annotated inputs,
$\shortcut(\Tt_U)$ is origin-equivalent to $\Tt_U$,
even when avoiding U-turns.
 
\begin{lemma}\label{lemma:shortcut1}
Let $w=u\otimes u^\lefttorightarrow\otimes u^\righttoleftarrow$ be
an arbitrary input annotated with left and right U-pairs.
Every successful run of $\Tt_U$ on $w$ is origin-equivalent 
to some successful run of $\shortcut(\Tt_U)$ on $w$ without lazy U-turns.
Conversely, every successful run of $\shortcut(\Tt_U)$ on $w$ is 
origin-equivalent to some successful run of $\Tt_U$ on $w$.
\end{lemma}

\begin{proof}
Consider an arbitrary run of $\Tt_U$  
on input $w=u\otimes u^\lefttorightarrow\otimes u^\righttoleftarrow$: 
\[
  \r ~=~
  (q_1,i_1) \act{b_1 \mid v_1} (q_2,i_2) \act{b_2 \mid v_2}
  (q_3,i_3) \cdots \act{b_m \mid v_m} (q_{m+1},i_{m+1}).
\]
The above run can also be seen as a run of $\shortcut(\Tt_U)$,
since for every transition of $\Tt_U$ outputting $v$ there is 
a similar transition of $\shortcut(\Tt_U)$, between the same
states and outputting the same word $v$.
We prove the claim by induction on the number of lazy U-turns in $\r$.
If $\rho$ has no lazy U-turn, then the claim follows trivially.
Now, for the inductive step, suppose that the factor that starts 
in $(q_k,i_k)$ and ends in $(q_h,i_h)$ is a maximal left lazy U-turn,
which thus occurs at position 
$i=i_k=i_h$
(the case of a right lazy U-turn case is symmetric).
By definition of U-turn, we know that $q_k$ is left-reading and $q_h$ is right-reading,
and hence the transitions that immediately precede and follow the U-turn
read the same input letter, i.e.~$w(i)=b_{k-1}=b_h$, and output respectively 
the words $v_{k-1}$ and $v_h$.
By construction, $\shortcut(\Tt_U)$ admits a transition $t$ 
that moves from configuration $(q_{k-1},i_{k-1})$ to configuration
$(q_h,i_h)$, reading $w(i)$ and producing $v_{k-1}\cdot v_h$ as output. 
We can replace the U-turn of $\r$ and the surrounding transitions 
with the latter transition $t$, thus obtaining an origin-equivalent 
run of $\shortcut(\Tt_U)$ with a smaller number of lazy U-turns. 
This proves the inductive step, and thus the first claim of the lemma.

For the second claim, consider an arbitrary run of $\shortcut(\Tt_U)$ on $w$:
 \[
  \r ~=~ 
  (q_1,i_1) \act{b_1 \mid v_1} (q_2,i_2) \act{b_2 \mid v_2}
  (q_3,i_3) \cdots \act{b_m \mid v_m} (q_{m+1},i_{m+1}).
\]
Consider a transition along $\r$, say
$t:~ (q_k,i_k) \act{b_k \mid v_k} (q_{k+1},i_{k+1})$.
By construction, there must be a sequence of states 
$r_1,r'_1,\dots,r_h,r'_h$ and directions $d_1,\dots,d_h$,
for some $h\ge 2$,
such that $q_k=r_1$, $q_{k+1}=r'_h$, $d_h=d$,
$(r_j,a,L_j,r'_j,d_j)\in\D$ and 
$(r'_j,r_{j+1})\in U^\lefttorightarrow\cup U^\righttoleftarrow$
for all $j$, 
and $v_k\in L_1\cdot L_2\cdots L_h$.
In particular, $\Tt_U$ admits some transitions of the form
$(q_k,i_k)=(r_1,j_1) \act{a\mid v'_1} (r'_1,j'_1)$,
$(r_2,j_2) \act{a\mid v'_2} (r'_2,j'_2)$, \dots,
$(r_h,j_h) \act{a\mid v'_h} (r'_h,j'_h)=(q_{k+1},i_{k+1})$,
with $j'_2=j_3$, \dots, $j'_{h-1}=j_h$, and $v_k=v'_1\cdot v'_2\cdots v'_h$. 
The latter transitions, interleaved with some U-turns
corresponding to the U-pairs 
$(r'_1,r_2),\dots,(r'_{h-1},r_h)\in U^\lefttorightarrow\cup U^\righttoleftarrow$,
form a valid run $\r'_t$ of $\Tt_U$ on $w$, which connects 
$(q_k,i_k)$ to $(q_{k+1},i_{k+1})$ and outputs $v_k$.
This means that in the run $\r$ we can safely replace 
the transition $t$ by the run $\r'_t$.
Doing this simultaneously for all transitions in $\r$ 
results in an origin-equivalent run of $\Tt_U$.
\end{proof}

The next step of the normalization consists of restricting 
the runs of $\shortcut(\Tt_U)$ so as to avoid any
lazy U-turn. This is done by simply forbidding the shortest 
possible lazy U-turns, namely, the transitions that output 
$\e$ and that remain on the same input position. On the one 
hand, since every lazy U-turn contains a transition of the 
previous form, forbidding this type of transitions results in forbidding
arbitrary lazy U-turns. On the other hand, thanks to Lemma~\ref{lemma:shortcut1}, 
this will not affect the semantics of $\shortcut(\Tt_U)$.
Formally, we construct from $\shortcut(\Tt_U)$ a new transducer 
$\norm(\Tt_U)$ by replacing every transition rule 
$(q,a,L,q',d)$ with $(q,a,L',q',d)$, where 
$L'$ is either $L$ or $L\setminus\{\e\}$, depending on whether
$q\in\Qleft\iff q'\in\Qleft$ or not.
We observe that $\norm(\Tt_U)$ has
the same set of states as the original transducer $\Tt$, and 
\PSPACE-constructible transitions w.r.t.~the size of $\Tt$.
The proof of the following result is straightforward and thus omitted.

\begin{lemma}\label{lemma:shortcut}
$\Tt_U$ and $\norm(\Tt_U)$ are origin-equivalent when 
restricted to correctly annotated inputs, i.e.:
$\sem{\Tt_U}_o \cap R \:=\: \sem{\norm(\Tt_U)}_o \cap R$,
where $R=\sem{\Uu}\times(\G\times\Nat)^*$.
\end{lemma}

\smallskip
We finally come to the last step of the reduction. This amounts to consider
two normalized transducers $\norm(\Tt_{1,U})$ and $\norm(\Tt_{2,U})$ that read
inputs annotated with the left/right U-pairs of both $\Tt_1$ and $\Tt_2$,
and replacing, in their transitions, the empty output by a special letter 
$\#\notin\G$. 
Formally, in the transition relation of $\norm(\Tt_{i,U})$, 
for $i=1,2$, we replace every tuple $(q,a,L,q',d)$, 
where $q$ is left-reading iff $q'$ is left-reading,
with the tuple $(q,a,L',q',d)$, where $L'$ is either 
$(L\setminus\{\e\})\cup\{\#\}$ or $L$, depending on whether $\e\in L$ or not.
The transducer obtained in this way is {\sl busy}, and is thus 
called $\busy(\Tt_{i,U})$.
Moreover, the states of $\busy(\Tt_{i,U})$ are the same as those of $\Tt_i$,
and its transitions are \PSPACE-constructible. The proposition below follows immediately from the previous arguments,
and reduces origin-containment between $\Tt_1$ and $\Tt_2$ to an origin-containment 
between $\busy(\Tt_{1,U})$ and $\busy(\Tt_{2,U})$, but relativized to correctly 
annotated inputs. 

\begin{proposition}\label{prop:lazy-equivalence}
Given two transducers $\Tt_1$ and $\Tt_2$, 
\[
  \sem{\Tt_1}_o ~\subseteq~ \sem{\Tt_2}_o
  \qquad\text{if and only if}\qquad
  \sem{\busy(\Tt_{1,U})}_o \:\cap\: R ~\subseteq~ \sem{\busy(\Tt_{2,U})}_o \:\cap\: R.
\]
where $R=\sem{\Uu'}\times(\G\times\Nat)^*$
and $\Uu'$ is an NFA that recognizes inputs 
annotated with left/right U-pairs of both $\Tt_1$ and $\Tt_2$.
\end{proposition}

It remains to show that, given the transducers $\Tt_1,\Tt_2$, 
there is a \PSPACE witness procedure for $\busy(\Tt_{1,U}),\busy(\Tt_{2,U})$:

\begin{proposition}\label{prop:witness}
Let $\Tt_1,\Tt_2$ be transducers with a total number $n$ of states,
and $\busy(\Tt_{i,U})=(Q_i,\hat{\S},\G,\D_i,I_i,F_i)$ for $i=1,2$
(the input alphabet $\hat{\S}$ is the same as for $\Uu'$). There is a
non-deterministic procedure $\Ww$ that works in polynomial space in
$n$ and returns on a given transition $t_1 = (q_1,a_1,q'_1,L_1,d_1)$
of $\busy(\Tt_{1,U})$ any set $X \subseteq \D_2$ of transitions of
$\busy(\Tt_{2,U})$ such that for some $v \in L_1$:
\[
  X ~=~ 
  \big\{t_2=(p_2,a_2,q_2,L_2,d_2) \in \D_2 \::\: 
        v \in L_2, \text{ and $t_2$ has {\sl same shape} as $t_1$} \big\}.
\]
\end{proposition}

\begin{proof}
Given transition $t_1 = (q_1,a,q'_1,L_1,d_1)$
  of $\busy(\Tt_{1,U})$, we first use the \PSPACE enumeration procedure 
  of the transitions of $\busy(\Tt_{2,U})$ to
  generate the set $Z$ of all
  $t_2=(q_2,a,q'_2,L_2,d_2) \in \D_2$ with the same shape as $t_1$. 
  Note that $Z$ has polynomial size, and
  its representation is polynomial as well. The procedure $\Ww$
  guesses first a subset
  $Z_0$ of $Z$. Then it guesses on-the-fly a word $v \in L_1$, and
  verifies that $v$ belongs to the output language of every $t_2 \in
  Z_0$, and to no output language of any $t_2 \in (Z \setminus Z_0)$. 
Recall that 
$\busy(\Tt_{1,U})$ and $\busy(\Tt_{2,U})$ 
have \PSPACE-constructible transitions w.r.t.~$n$, 
and thus, by definition, the output languages 
are represented by NFAs of size polynomial in $n$.
Basically $\Ww$ checks a condition of the form 
$v \in \bigcap_{i} \Dd_i \:\setminus\: \bigcup_j \Dd'_j$, 
for polynomially many NFAs $\Dd_i,\Dd'_j$ of size polynomial in $n$,
which can be done in \PSPACE.
  If successful, then $\Ww$ returns the set $Z_0$.
\end{proof}

We can now conclude with the proof of our main result, which we recall here:

\begin{reptheorem}{thm:main}
Containment and equivalence of two-way transducers 
under the origin semantics is \PSPACE-complete.
\end{reptheorem}

\begin{proof}
The lower bound follows from a straightforward reduction 
from the classical equivalence problem of NFA.
For the upper bound, in view of Proposition \ref{prop:lazy-equivalence},
we can consider origin containment between the transducers 
$\busy(\Tt_{1,U})$ and $\busy(\Tt_{2,U})$, 
obtained from $\Tt_1$ and $\Tt_2$, respectively.
We recall that $\busy(\Tt_{i,U})$ has at most $n_i=|\Tt_i|$ 
states and \PSPACE-constructible transitions w.r.t.~$n_i$,
for $i=1,2$.
We then apply Proposition \ref{prop:from-containment-to-witness} 
and Lemma \ref{lemma:from-witness-to-nfa} to reduce the 
containment problem to an emptiness problem for the intersection 
of two \PSPACE-constructible NFA, i.e.~$\Bb$ and $\Uu'$.
We observe that $\Uu'$ is basically the product of two NFA 
obtained from Lemma \ref{lemma:annotations} by letting 
$\Tt=\Tt_i$, once for $i=1$ and once for $i=2$.
Finally, we use the same arguments as in the proof of 
Corollary \ref{cor:busy-containment} to conclude that 
the latter emptiness problem is decidable in \PSPACE w.r.t.~$n=n_1+n_2$. 
\end{proof}

 \section{Containment modulo resynchronization}\label{sec:resynch}

In this section we aim at generalizing the equivalence and containment 
problems for transducers with origins. The goal is to compare 
the origin semantics of any two transducers up to  `distortions', that is, 
 differences
in the origin tagging each position of the output.

Recall that
$\sem{\Tt}_o$ is the set of synchronized pairs $\oelement=(u,\nu)$, 
where $u\in\Sigma^*$ is a possible input for the transducer $\Tt$ and 
$\nu\in(\Gamma\times\Nat)^*$ is an output (tagged with origins)
produced by a successful run of $\Tt$ on $u$. 
Given a pair $\oelement=(u,\nu)\in\sem{\Tt}_o$, we denote 
by $\In(\oelement)$, $\Out(\oelement)$, and $\Orig(\oelement)$ respectively
the input word $u$, the output word obtained by projecting $\nu$ onto the 
finite alphabet $\Gamma$, and the sequence of input positions obtained by 
projecting $\nu$ onto $\Nat$.
This notation is particularly convenient for describing resynchronizations,
that is, relations between synchronized pairs.
Following prior terminology from~\cite{FiliotJLW16},
we call \emph{resynchronization} any relation $R$ between 
synchronized pairs that preserves input and output words,
but can modify the origins, namely, such that 
$(\oelement,\oelement')\in R$ implies 
$\In(\oelement)=\In(\oelement')$ and $\Out(\oelement)=\Out(\oelement')$.

The \emph{containment problem modulo a resynchronization $R$}~\cite{FiliotJLW16}
is the problem of deciding, given two transducers $\Tt_1,\Tt_2$, 
if for every $\oelement'\in\sem{\Tt_1}_o$, there is $\oelement\in\sem{\Tt_2}_o$
such that $(\oelement,\oelement')\in R$, 
or in other words if every synchronized pair of $\Tt_1$
can be seen as a distortion of a synchronized pair of $\Tt_2$.
In this case we write for short $\Tt_1\subseteq_R \Tt_2$.
We remark that, despite the name `containment' and the notation,
the relation $\subseteq_R$ is not necessarily transitive, 
in the sense that it may happen that $\Tt_1\subseteq_R \Tt_2$ and
$\Tt_2\subseteq_R \Tt_3$, but $\Tt_1\not\subseteq_R \Tt_3$.

We propose a class of resynchronizations that can be described in
monadic second-order logic (\emph{MSO}).  The spirit is that, as
synchronized pairs are (special) graphs, graph transformations à la
Courcelle and Engelfriet~\cite{CourBook12} are an adequate tool to
define resynchronizers. However, we cannot directly use MSO logic over
origin graphs, since this would result in an undecidable containment
problem. Our MSO resynchronizers will be able to talk about (regular)
properties of the input word and the output word, and say how origins
are distorted. We will show that containment modulo MSO
resynchronizations is decidable, assuming a (decidable) restriction
on the change of origins.

\begin{definition}\label{def:resynch}
An \emph{MSO resynchronizer} $R$ is a tuple $(\alpha,\beta,\gamma,\delta)$, where
\begin{itemize}
  \item $\alpha(\bar I)$ is an MSO formula over the signature of the input word,
        and has some free monadic variables $\bar I=(I_1,\dots,I_m)$, 
        called \emph{input parameters},
  \item $\beta(\bar O)$ is an MSO formula over the signature of the output word,
        and has some free monadic variables $\bar O=(O_1,\dots,O_n)$,
        called \emph{output parameters},
  \item $\gamma$ is a function that maps any element $\otype\in\Gamma\times\bbB^n$,
        with $\bbB=\{0,1\}$,
        to an MSO formula $\gamma(\otype)(\bar I,y,z)$ over the input signature
        that has a tuple of free monadic variables $\bar I$ (input parameters) and 
        two free first-order variables $y,z$
        (called \emph{source} and \emph{target}, respectively),
  \item $\delta$ is a function that maps any pair of elements $\otype,\otype'\in\Gamma\times\bbB^n$,
        with $\bbB=\{0,1\}$,
        to an MSO formula $\delta(\otype,\otype')(\bar I,z,z')$ over the input signature
        that has a tuple of free monadic variables $\bar I$ (input parameters)
        and two free first-order variables $z,z'$ (called \emph{targets}).
\end{itemize}
\end{definition}

\noindent
The input and output parameters $\bar I$, $\bar O$ that appear 
in the formulas of an MSO resynchronizer play the same role as the parameters of an 
NMSO-transduction~\cite{eh01}. They allow to express regular
properties of the input and output word, respectively, through the first two formulas, 
$\alpha(\bar I)$ and $\beta(\bar O)$. 
The other two formulas are used to describe how the origin function is
transformed. They depend on the input and output parameters, in
particular, on the ``type'' of the positions of the output word, as
defined next.

Given an output word $v=\Out(\oelement)$ and an interpretation 
$\bar O=O_1,\dots,O_n\subseteq\dom(v)$ for the output parameters, 
let us call \emph{output-type} of a position $x\in\dom(v)$ 
the element $\otype=(v(x),b_1,\dots,b_n) \in \Gamma\times\bbB^n$, 
where each $b_i$ is either $1$ or $0$ 
depending on whether $x\in O_i$ or not.
Based on the output-type $\otype$ of $x$, 
the formula $\gamma(\otype)(\bar I,y,z)$ describes how the origin of 
$x$ is redirected from a source $y$ to a target $z$ in the input word.
Similarly, $\delta(\otype,\otype')(\bar I,z,z')$ constraints 
the origins $z,z'$ of two consecutive output positions $x,x+1$, 
based on their output-types $\otype$ and $\otype'$. This is formalized below.

\begin{definition}\label{def:resynch-semantics}
An MSO resynchronizer $R=(\alpha,\beta,\gamma,\delta)$ induces the 
resynchronization $\sem{R}$ defined by $(\oelement,\oelement') \in \sem{R}$ 
if and only if $\In(\oelement)=\In(\oelement')$, 
$\Out(\oelement)=\Out(\oelement')$,  
and there are $\bar I=I_1,\dots,I_m\subseteq\dom(\In(\oelement))$ 
and $\bar O=O_1,\dots,O_n\subseteq\dom(\Out(\oelement))$ such that:
\begin{itemize}
  \item $(\In(\oelement),\bar I)\sat\alpha$ (or equally, $(\In(\oelement'),\bar I)\sat\alpha$),
  \item $(\Out(\oelement),\bar O)\sat\beta$ (or equally, $(\Out(\oelement'),\bar O)\sat\beta$),
  \item for all $x\in\dom(\Out(\oelement))$ with output-type $\otype$, 
                if $\Orig(\oelement)(x)=y$ and $\Orig(\oelement')(x)=z$, 
        then $(\In(\oelement),\bar I,y,z)\sat\gamma(\otype)$,
  \item for all pairs of consecutive positions $x$ and $x+1$ in $\Out(\oelement)$ 
        with output-types $\otype$ and $\otype'$, respectively,
                if $\Orig(\oelement')(x)=z$ and $\Orig(\oelement')(x+1)=z'$, 
        then $(\In(\oelement),\bar I,z,z')\sat\delta(\otype,\otype')$.
\end{itemize}
\end{definition}

\begin{example}\label{ex:resync}
We provide a few examples of MSO resynchronizers:
\begin{enumerate}
\smallskip
\item The most unconstrained resynchronization, called \emph{universal resynchronization},
groups any two synchronized pairs with the same input and output strings.
This is readily defined by an MSO resynchronizer 
$R_{\mathit{univ}}=(\alpha,\beta,\gamma,\delta)$ without parameters,
where $\alpha=\beta=\gamma(\otype)(y,z)=\delta(\otype,\otype')(z,z')=\true$ 
for all $\otype,\otype'\in\G$. 
\smallskip
\item 
\mywraptextjustified{4cm}{Consider a resynchronization that displaces the origin in a
synchronized pair by one position to the left or to the right.
An instance of this resynchronization is shown to the right, where solid arrows 
represent origins in the initial synchronized pair,}\mywrapfig{4cm}
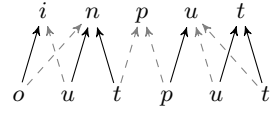
{\vspace{-2mm}
\begin{tikzpicture}[scale=0.65,baseline]
  \draw (0,1.7) node [hidden] (I1) {\small $i$};
  \draw (1,1.7) node [hidden] (I2) {\small $n$};
  \draw (2,1.7) node [hidden] (I3) {\small $p$};
  \draw (3,1.7) node [hidden] (I4) {\small $u$};
  \draw (4,1.7) node [hidden] (I5) {\small $t$};
  
  \draw (-0.5,0) node [hidden] (O0) {\small $o$};
  \draw (0.5,0) node [hidden] (O1) {\small $u$};
  \draw (1.5,0) node [hidden] (O2) {\small $t$};
  \draw (2.5,0) node [hidden] (O3) {\small $p$};
  \draw (3.5,0) node [hidden] (O4) {\small $u$};
  \draw (4.5,0) node [hidden] (O5) {\small $t$};
  
  \draw (O0) edge [arrow, gray, dashed] (I2);
  \draw (O1) edge [arrow, gray, dashed] (I1);
  \draw (O2) edge [arrow, gray, dashed] (I3);
  \draw (O3) edge [arrow, gray, dashed] (I3);
  \draw (O4) edge [arrow, gray, dashed] (I4);
  \draw (O5) edge [arrow, gray, dashed] (I4);

  \draw (O0) edge [arrow] (I1);
  \draw (O1) edge [arrow] (I2);
  \draw (O2) edge [arrow] (I2);
  \draw (O3) edge [arrow] (I4);
  \draw (O4) edge [arrow] (I5);
  \draw (O5) edge [arrow] (I5);
\end{tikzpicture}
\hspace{-5mm} }
and dashed arrows represent origins in the modified synchronized pair.
This transformation can be defined by a parameterless MSO resynchronizer 
$R_{\pm1}=(\alpha,\beta,\gamma,\delta)$, 
where $\alpha=\beta=\true$, $\gamma(\otype)(y,z)=(z=y-1) \vee (z=y+1)$, 
and $\delta(\otype,\otype')(z,z')=\true$ for all $\otype,\otype'\in\G$.
Note that  the reflexive and transitive closure of $\sem{R_{\pm1}}$
gives the universal resynchronization $\sem{R_{\mathit{univ}}}$.
\smallskip
\item 
\mywraptextjustified{4.5cm}{We give a variant of the previous resynchronizer $R_{\pm1}$,
where the direction of movement of the origin is controlled by a
property of the output position, e.g.~the parity of the number of 
$b$'s that precede that position in the output.
We see an instance of the resynchronization to the right (as}\mywrapfig{4.5cm}
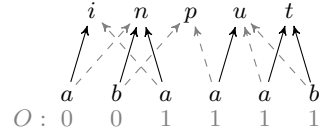
{\vspace{-2mm}
\begin{tikzpicture}[scale=0.65]
  \draw (0,1.7) node [hidden] (I1) {\small $i$};
  \draw (1,1.7) node [hidden] (I2) {\small $n$};
  \draw (2,1.7) node [hidden] (I3) {\small $p$};
  \draw (3,1.7) node [hidden] (I4) {\small $u$};
  \draw (4,1.7) node [hidden] (I5) {\small $t$};
  
  \draw (-0.5,0) node [hidden] (O0) {\small $a$};
  \draw (0.5,0) node [hidden] (O1) {\small $b$};
  \draw (1.5,0) node [hidden] (O2) {\small $a$};
  \draw (2.5,0) node [hidden] (O3) {\small $a$};
  \draw (3.5,0) node [hidden] (O4) {\small $a$};
  \draw (4.5,0) node [hidden] (O5) {\small $b$};

  \draw (-1.25,-0.5) node [hidden, gray] {\small $O:$};
  \draw (-0.5,-0.5) node [hidden, gray] {\small $0$};
  \draw (0.5,-0.5) node [hidden, gray] {\small $0$};
  \draw (1.5,-0.5) node [hidden, gray] {\small $1$};
  \draw (2.5,-0.5) node [hidden, gray] {\small $1$};
  \draw (3.5,-0.5) node [hidden, gray] {\small $1$};
  \draw (4.5,-0.5) node [hidden, gray] {\small $1$};
   
  \draw (O0) edge [arrow, gray, dashed] (I2);
  \draw (O1) edge [arrow, gray, dashed] (I3);
  \draw (O2) edge [arrow, gray, dashed] (I1);
  \draw (O3) edge [arrow, gray, dashed] (I3);
  \draw (O4) edge [arrow, gray, dashed] (I4);
  \draw (O5) edge [arrow, gray, dashed] (I4);

  \draw (O0) edge [arrow] (I1);
  \draw (O1) edge [arrow] (I2);
  \draw (O2) edge [arrow] (I2);
  \draw (O3) edge [arrow] (I4);
  \draw (O4) edge [arrow] (I5);
  \draw (O5) edge [arrow] (I5);
\end{tikzpicture}
\hspace{-5mm} }
usual, solid and dashed arrows represent initial and modified origins).
The transformation can be defined by an MSO resynchronizer 
with a single output parameter $O$ that encodes
the parity condition 
(an interpretation of $O$ is shown in the figure 
as an annotation of the output over $\mathbb{B}=\{0,1\}$). 
Formally, we let $R_{\pm1}'=(\alpha,\beta,\gamma,\delta)$, where
$\alpha=\true$, 
$\beta(O)=\forall x ~ 
          \big(b(x) \rightarrow (x\in O \leftrightarrow x+1\not\in O)\big) \wedge 
          \big(\neg b(x) \rightarrow (x\in O \leftrightarrow x+1\in O)\big) \wedge
          (x=\mathit{first} \rightarrow x\in O))$,
and $\gamma$ and $\delta$ are defined on the basis of the
output-types $\otype,\otype'\in\G\times\mathbb{B}$, as follows:
$\gamma(\otype)(y,z)$ enforces either $z=y+1$ or $z=y-1$ depending
on whether $\otype\in\G\times\{0\}$ or $\otype\in\G\times\{1\}$,
and $\delta(\otype,\otype')(z,z')=\true$.
\smallskip
\item 
\mywraptextjustified{5cm}{Let us now consider a resynchronization that does not modify the origins, 
but only constrain them so as to obtain a `regular' subset of 
synchronization pairs. Here the allowed synchronization pairs are 
those that correspond to the process of applying a rational substitution
$f:\Sigma\rightarrow 2^{\Gamma^*}$ to an input over $\Sigma$. 
The figure to the right describes a possible synchronized pair for $f$
defined by}\mywrapfig{5cm}
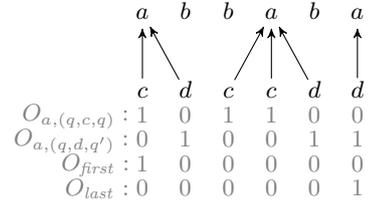
{\vspace{-2mm}
\begin{tikzpicture}[scale=0.65,xscale=0.87]
  \draw (0,1.6) node [hidden] (I1) {\small $a$};
  \draw (1,1.6) node [hidden] (I2) {\small $b$};
  \draw (2,1.6) node [hidden] (I3) {\small $b$};
  \draw (3,1.6) node [hidden] (I4) {\small $a$};
  \draw (4,1.6) node [hidden] (I5) {\small $b$};
  \draw (5,1.6) node [hidden] (I6) {\small $a$};
  
  \draw (0,0) node [hidden] (O1) {\small $c$};
  \draw (1,0) node [hidden] (O2) {\small $d$};
  \draw (2,0) node [hidden] (O3) {\small $c$};
  \draw (3,0) node [hidden] (O4) {\small $c$};
  \draw (4,0) node [hidden] (O5) {\small $d$};
  \draw (5,0) node [hidden] (O6) {\small $d$};

  \draw (-1.55,-0.5) node [hidden, gray] {\small $O_{a,(q,c,q)}:$};
  \draw (0,-0.5) node [hidden, gray] {\small $1$};
  \draw (1,-0.5) node [hidden, gray] {\small $0$};
  \draw (2,-0.5) node [hidden, gray] {\small $1$};
  \draw (3,-0.5) node [hidden, gray] {\small $1$};
  \draw (4,-0.5) node [hidden, gray] {\small $0$};
  \draw (5,-0.5) node [hidden, gray] {\small $0$};
   
  \draw (-1.65,-1) node [hidden, gray] {\small $O_{a,(q,d,q')}:$};
  \draw (0,-1) node [hidden, gray] {\small $0$};
  \draw (1,-1) node [hidden, gray] {\small $1$};
  \draw (2,-1) node [hidden, gray] {\small $0$};
  \draw (3,-1) node [hidden, gray] {\small $0$};
  \draw (4,-1) node [hidden, gray] {\small $1$};
  \draw (5,-1) node [hidden, gray] {\small $1$};
   
  \draw (-1.09,-1.5) node [hidden, gray] {\small $O_{\mathit{first}}:$};
  \draw (0,-1.5) node [hidden, gray] {\small $1$};
  \draw (1,-1.5) node [hidden, gray] {\small $0$};
  \draw (2,-1.5) node [hidden, gray] {\small $0$};
  \draw (3,-1.5) node [hidden, gray] {\small $0$};
  \draw (4,-1.5) node [hidden, gray] {\small $0$};
  \draw (5,-1.5) node [hidden, gray] {\small $0$};
   
  \draw (-1.05,-2) node [hidden, gray] {\small $O_{\mathit{last}}:$};
  \draw (0,-2) node [hidden, gray] {\small $0$};
  \draw (1,-2) node [hidden, gray] {\small $0$};
  \draw (2,-2) node [hidden, gray] {\small $0$};
  \draw (3,-2) node [hidden, gray] {\small $0$};
  \draw (4,-2) node [hidden, gray] {\small $0$};
  \draw (5,-2) node [hidden, gray] {\small $1$};

  \draw (O1) edge [arrow] (I1);
  \draw (O2) edge [arrow] (I1);
  \draw (O3) edge [arrow] (I4);
  \draw (O4) edge [arrow] (I4);
  \draw (O5) edge [arrow] (I4);
  \draw (O6) edge [arrow] (I6);
\end{tikzpicture}
\hspace{-5mm} }
$f(a)=c^*d$ and $f(b)=\varepsilon$.
Below, we show how to define such a resynchronization
for an arbitrary rational substitution $f$.
We fix, for each letter $a\in\Sigma$, 
an NFA $\Aa_a$ that recognizes the regular language $f(a)$.
We then define an MSO resynchronizer $R_f$ that uses 
one output parameter $O_{a,t}$ for every letter $a\in\Sigma$
and every transition $t$ of $\Aa_a$, plus two additional output 
parameters $O_{\mathit{first}}$ and $O_{\mathit{last}}$.
By a slight abuse of notation, given the output-type $\otype$ of a position $x$,
we write $\otype[a,t]=1$ whenever 
$x\in O_{a,t}$, and similarly for $\otype[\mathit{first}]$ and $\otype[\mathit{last}]$.
The first formula $\alpha$ of the resynchronizer holds 
vacuously, as we have no restriction on the input. 
The second formula $\beta$ requires that 
the parameters $O_{a,t}$ form a partition of the output domain
in such a way that if a position $x$ is labeled by a letter 
$c\in\G$ and $x\in O_{a,t}$, then $t$ is a $c$-labeled transition
of $\Aa_a$. In addition, $\beta$ requires that the parameters
$O_{\mathit{first}}$ and $O_{\mathit{last}}$ are singletons 
consisting of the first and the last output position, respectively.
The third component $\gamma$ of the resynchronizer is defined by
$\gamma(\otype)(y,z)= a(y) \wedge (y=z)$ whenever
$\otype[a,t]=1$ (the origin is not modified, 
and the transition annotating the output position must belong to the correct NFA). 
The last component $\delta$ restricts further the parameters and the origins 
for consecutive output positions, so as to simulate successful runs of the NFA.
Formally, $\delta(\otype,\otype')(z,z')$ enforces the following constraints:
\begin{itemize}
  \item if $\otype[a,t]=1$, $\otype'[a,t']=1$, and $z=z'$, 
        then the target state of $t$ coincide with the source state of $t'$,
        namely, $t t'$ forms a factor of a run of $\Aa_a$,
  \item if $\otype[a,t]=1$, $\otype'[a',t']=1$, and $z<z'$, 
        then the target state of $t$ must be final, the source state of $t'$ must be initial,
        and every input letter strictly between $z$ and $z'$ is mapped via $f$ to a language 
        that contains $\varepsilon$,
  \item if $\otype[\mathit{first}]=1$ and $\otype[a,t]$, then the source state of $t$ must 
        be initial, and every input letter strictly before $z$ is mapped via $f$ to a 
        language that contains $\varepsilon$, 
  \item if $\otype'[\mathit{last}]=1$ and $\otype'[a,t]$, then the target state of $t$ must 
        be final, and every input letter strictly after $z'$ is mapped via $f$ to a 
        language that contains $\varepsilon$.
\end{itemize}
\smallskip
\item
\mywraptextjustified{4cm}{We conclude the list of examples with a resynchronization
that moves the origin of an arbitrarily long output over 
an arbitrarily long distance. 
This resynchronization contains the pairs $(\oelement,\oelement')$,
where $\oelement$ (resp.~$\oelement'$) maps every output position
to the}\mywrapfig{4cm}
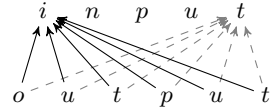
{\vspace{-2mm}
\begin{tikzpicture}[scale=0.65,baseline]
  \draw (0,1.7) node [hidden] (I1) {\small $i$};
  \draw (1,1.7) node [hidden] (I2) {\small $n$};
  \draw (2,1.7) node [hidden] (I3) {\small $p$};
  \draw (3,1.7) node [hidden] (I4) {\small $u$};
  \draw (4,1.7) node [hidden] (I5) {\small $t$};
  
  \draw (-0.5,0) node [hidden] (O0) {\small $o$};
  \draw (0.5,0) node [hidden] (O1) {\small $u$};
  \draw (1.5,0) node [hidden] (O2) {\small $t$};
  \draw (2.5,0) node [hidden] (O3) {\small $p$};
  \draw (3.5,0) node [hidden] (O4) {\small $u$};
  \draw (4.5,0) node [hidden] (O5) {\small $t$};
  
  \draw (O0) edge [arrow, gray, dashed] (I5);
  \draw (O1) edge [arrow, gray, dashed] (I5);
  \draw (O2) edge [arrow, gray, dashed] (I5);
  \draw (O3) edge [arrow, gray, dashed] (I5);
  \draw (O4) edge [arrow, gray, dashed] (I5);
  \draw (O5) edge [arrow, gray, dashed] (I5);

  \draw (O0) edge [arrow] (I1);
  \draw (O1) edge [arrow] (I1);
  \draw (O2) edge [arrow] (I1);
  \draw (O3) edge [arrow] (I1);
  \draw (O4) edge [arrow] (I1);
  \draw (O5) edge [arrow] (I1);
\end{tikzpicture}
\hspace{-5mm} }
first (resp.~last) input position, as shown in the figure.
This is defined by the MSO resynchronizer
$R_{\mathit{1st-to-last}}=(\alpha,\beta,\gamma,\delta)$, 
where $\alpha=\beta=\delta(\otype,\otype')(z,z')=\true$ and 
$\gamma(\otype)(y,z)=(y=\mathit{first}) \wedge (z=\mathit{last})$, 
for all $\otype,\otype'\in\G$.
Note that the resynchronization $\sem{R_{\mathit{1st-to-last}}}$ 
is also `one-way', in the sense that it contains only synchronized 
pairs that are admissible outcomes of runs of one-way transducers. 
However, $\sem{R_{\mathit{1st-to-last}}}$ is not captured
by the formalism of one-way {\sl rational} resynchronizers from \cite{FiliotJLW16}.
\end{enumerate}
\end{example}

\noindent
Recall the Example \ref{ex:resync}.1 above, which defines the
universal resynchronization $\sem{R_{\mathit{univ}}}$. 
The containment problem modulo $\sem{R_{\mathit{univ}}}$
boils down to testing classical containment between transducers
{\sl without origins}, which is known to be undecidable \cite{gri68}.
Based on this, it is clear that in order to compare effectively 
transducers modulo resynchronizations, we need to restrict further 
our notion of MSO resynchronizer. Intuitively, what makes the 
containment problem modulo resynchronization undecidable is the 
possibility of redirecting many sources to the same target. 
This possibility is explicitly forbidden in the definition below.
Also observe that, since $\sem{R_{\mathit{univ}}}$ is the reflexive 
and transitive closure of $\sem{R_{\pm1}}$, we cannot take our 
resynchronizations to be equivalence relations.

\begin{definition}\label{def:regular-resync}
An MSO resynchronizer $(\alpha,\beta,\gamma,\delta)$ is \emph{$k$-bounded}
if for all inputs $u$, parameters $\bar I=I_1,\dots,I_m\subseteq\dom(u)$, 
output-types $\otype\in\Gamma\times\bbB^n$, and targets $z\in\dom(u)$,
there are at most $k$ distinct sources $y_1,\dots,y_k\in\dom(u)$ 
such that $(u,\bar I,y_i,z)\sat\gamma(\otype)$ for all $i=1,\dots,k$.
An MSO resynchronizer is \emph{bounded} if it is $k$-bounded for some $k$. 
\end{definition}

Note that all resynchronizers from Example \ref{ex:resync} 
but $R_{\mathit{univ}}$ are bounded. For instance, $R_{\mathit{1st-to-last}}$
is $1$-bounded and $R_{\pm1}$ is $2$-bounded.

\begin{proposition}\label{prop:boundedness}
t is decidable to know whether a given MSO resynchronizer is bounded.
\end{proposition}

\begin{proof}
Let $R=(\alpha,\beta,\gamma,\delta)$ be a resynchronizer as above. 
Since we can work with words over $\S'=\S \times  \bbB^{m+1}$,
where the annotation over $\bbB^{m+1}$ encodes the interpretation of 
the input parameters $\bar I$ and the target $z$ of $\gamma(\bar I,y,z)$,
verifying whether $R$ is bounded amounts to verify whether 
an MSO formula $\tilde\gamma(y)$ with one free variable satisfies 
the following property:

\begin{quote}
\emph{There is some $k$ such that for every word $w \in {\S'}^*$, 
there are at most $k$ positions $y \in\dom(w)$ such that $w,y \sat \tilde\gamma$.}
\end{quote}

\noindent
To check the above property we assume that $\tilde\gamma$ 
is given by a deterministic finite automaton (DFA) $\Aa$. 
The DFA processes words with a single marked position, 
i.e.~words over $\S' \times \bbB$. 

From $\Aa=(Q,\S' \times \bbB,\D,q_0,F)$ we build an NFA $\Bb$ 
that recognizes the projection of $\Aa$ over $\Sigma'$. 
Formally, the NFA $\Bb$ has $\S'$ as input alphabet, and two 
copies of $Q$ as set of states, i.e.~$Q'=Q \uplus \hat Q$.
The two copies of $Q$ have transitions of the form $(q,a,q')$ 
and $(\hat q,a,\hat q')$, for any $\big(q,(a,0),q')\big)\in\D$.
In addition, if $\big(q,(a,1),q'\big)\in\D$,
then we add the transition $(q,a,\hat q')$ that moves from $Q$ to $\hat Q$.

An NFA is said to be \emph{$k$-ambiguous} if it admits at most $k$ 
successful runs on every given input. 
It is \emph{finitely ambiguous} if it is $k$-ambiguous for some $k$.
From the above constructions it follows that $\Bb$ is finitely ambiguous 
if and only if the MSO resynchronizer is bounded.
To conclude, we recall that one can decide in polynomial time whether
a given NFA is finitely ambiguous~\cite{WeberS91}. 
\end{proof}

The goal is to reduce the problem of containment modulo a bounded MSO resynchronizer 
to a standard containment problem in the origin semantics. For this, the
natural approach would be to show that transducers are effectively closed under
bounded MSO resynchronizers. 
However, this closure property cannot be proven in full generality,
because of the (input) parameters that occur in the definition of
resynchronizers. More precisely, {\sl two-way} transducers cannot guess parameters
in a consistent way (different guesses could be made at different
visits of the same input position). We could show the closure if we adopted 
a slightly more powerful notion of transducer, with so-called 
\emph{common guess}~\cite{BDGPicalp17}.
Here we prefer to work with classical two-way transducers and explicitly 
deal with the parameters. Despite the different terminology,
the principle is the same: parameters are guessed beforehand and 
accessed by the two-way transducer as explicit annotations of the input. 
Given a transducer $\Tt$ over an expanded input alphabet $\Sigma\times\Sigma'$
and an NFA $\Aa$ over $\Sigma\times\Sigma'$, we let
\[
  \sem{\Tt}_o\prj[\Aa]{\Sigma} 
  ~=~ \big\{ (u,\nu) \in \S^*\times(\G\times\Nat)^* \::\: 
             \exists ~ u'\in{\Sigma'}^{|u|}  ~~ u\otimes u'\in \sem{\Aa}, ~
                                                (u\otimes u',\nu)\in\sem{\Tt}_o \big\}.
\]
In other words, $\sem{\Tt}_o\prj[\Aa]{\Sigma}$ is
obtained from $\sem{\Tt}_o$ by restricting the inputs of $\Tt$ via
$\Aa$, and then projecting them on $\S$. 

The following result is the key to reduce containment modulo bounded
MSO resynchronizers to containment in the origin semantics.

\begin{theorem}\label{thm:resync}
Given a bounded MSO resynchronizer $R$ with $m$ input parameters, 
a transducer $\Tt$ with input alphabet $\Sigma\times\Sigma'$
and an NFA $\Aa$ over $\Sigma\times\Sigma'$, one can 
construct a transducer $\Tt'$ with input alphabet $\Sigma\times\Sigma'\times\bbB^m$
and an NFA $\Aa'$ over $\Sigma\times\Sigma'\times\bbB^m$ such that
\[
  \sem{\Tt'}_o\prj[\Aa']{\Sigma} ~=~ 
  \big\{ \oelement' ~:~  (\oelement,\oelement')\in\sem{R}
  \text{ for some } \oelement\in\sem{\Tt}_o\prj[\Aa]{\Sigma}
         \big\}. 
\]
Moreover, if $R$ is fixed, 
$\Tt$ has $n$ states and \PSPACE-constructible transitions w.r.t.~$n$,
and $\Aa$ is \PSPACE-constructible w.r.t.~$n$, 
then $\Tt'$ and $\Aa'$ have similar properties, namely,
$\Tt'$ has a number of states polynomial in $n$ and
\PSPACE-constructible transitions w.r.t.~$n$,
and $\Aa'$ is \PSPACE-constructible w.r.t.~$n$.
\end{theorem}

\begin{proof}
To prove the claim, it is convenient to assume that $R$ has no
input nor output parameters. If this were not the case, we could 
modify $\Tt$ in such a way that it reads inputs over $\S\times\S'\times\bbB^m$,
exposing a valuation of the parameters $\bar I$, and produces 
outputs over $\G\times\bbB^n$, exposing a valuation $\bar O$. 
We could then apply the constructions that follow, and finally 
modify the resulting transducer $\Tt'$ by projecting away the 
input and output annotations.
We observe that the projection operation on the output 
is easier and can be implemented directly at the level of the 
transitions of $\Tt'$, while the projection of the input 
requires the use of the notation $\sem{\Tt'}_o\prj[\Aa']{\Sigma}$,
for the reasons that we discussed earlier. In both cases the 
complexity bounds are preserved.

Let $R=(\alpha,\beta,\gamma,\delta)$ a bounded MSO resynchronizer 
with no input/output parameters. Since there are no existentially 
quantified parameters, $\sem{R}$ can be seen as the relational 
composition of four different resynchronizations, as induced by 
the formulas of $R$. Formally, we have 
$\sem{R}=\sem{R_\alpha} \circ \sem{R_\beta} \circ \sem{R_\gamma} \circ \sem{R_\delta}$, 
where 
$R_\alpha=(\alpha,\true,\gamma_{\mathit{id}},\delta_{\true})$, 
$R_\beta=(\true,\beta,\gamma_{\mathit{id}},\delta_{\true})$, 
$R_\gamma=(\true,\true,\gamma,\delta_{\true})$,  
$R_\delta=(\true,\true,\gamma_{\mathit{id}},\delta)$,
$\gamma_{\mathit{id}}(\otype)(y,z)=(y=z)$ 
and $\delta_{\true}(\otype,\otype')(z,z')=\true$ 
for all output-types $\otype,\otype'\in\Gamma$.
This means that, to prove the claim, it suffices to consider 
only one resynchronizer at a time among 
$R_\alpha$, $R_\beta$, $R_\gamma$, $R_\delta$.

\smallskip
Let us begin with the resynchronizer 
$R_\alpha=(\alpha,\true,\gamma_{\mathit{id}},\delta_{\true})$.
In this case the origins are not modified: the only effect
of $R_\alpha$ is to restrict the set of possible inputs. 
Formally, we need to check that an input $u\in\Sigma^*$, 
not only is accepted by the NFA $\Aa$, but also satisfies the formula $\alpha$.
This can be checked by an NFA $\Aa'$ that is obtained from intersecting 
$\Aa$ with an NFA equivalent to $\alpha$.
In this case, the transducer $\Tt'$ is the same as $\Tt$,
and the construction of $\Aa'$ takes time polynomial in $|\Aa|$, 
assuming that $\alpha$ is fixed. In particular this satisfies
the complexity bounds given in the second claim of the theorem.

\smallskip
Let us now consider the resynchronizer
$R_\beta=(\true,\beta,\gamma_{\mathit{id}},\delta_{\true})$, 
which restricts the set of possible outputs.
In this case, since no further restriction is enforced 
on the input, we can take the NFA $\Aa'$ to be the same as $\Aa$. 
To construct $\Tt'$, we proceed as follows.
We first construct an NFA $\Bb$ equivalent to $\beta$. 
For every pair of states $q,q'$ of $\Bb$, let $L_{q,q'}$ 
be the language of words that induce runs of $\Bb$ from $q$ to $q'$.
We use the latter languages to modify $\Tt$ and simulate 
a successful run of $\Bb$ on the output. 
Formally, the control states of $\Tt$ are paired with the control 
states of $\Bb$, and the new transition rules are of the form
$(p,q) \act{a|L \cap L_{q,q'}} (p',q')$, whenever $p \act{a|L} p'$ 
is a transition rule of $\Tt$ and $L \cap L_{q,q'} \neq \emptyset$. 
The initial and final states of $\Tt'$
are defined accordingly.
As before, the construction of $\Tt'$ takes polynomial time in $|\Tt|$,
assuming that the formula $\beta$ is fixed.

\smallskip
We now consider the most interesting case, that of 
$R_\gamma=(\true,\true,\gamma,\delta_{\true})$, which modifies the
origins.
The rough idea here is that $\Tt'$ has to simulate an arbitrary run of $\Tt$, 
by displacing the origins of any output letter with type $\otype$ from a 
source $y$ to a target $z$, as indicated by the formula $\gamma(\otype)(x,y)$. 
Since a factor of an output of $\Tt$ that originates at the same input position
can be broken up into multiple sub-factors with origins at different positions,
here it is convenient to assume that $\Tt$ outputs at most a single letter at 
each transition.
This assumption can be made without loss of generality, since
we can reproduce any longer word $v$ that is output by some transition $(q,i) \act{a|L} (q',i')$ 
letter by letter, with several transitions that move back and forth around position $i$.

The idea of the construction is as follows.
Whenever $\Tt$ outputs a letter $b$ with origin in $y$, 
$\Tt'$ non-deterministically moves to some position $z$ that, together with $y$, 
satisfies the formula $\gamma(b)$ (note that $b$ is also the output-type of the 
produced letter).
Then $\Tt'$ produces the same output $b$ as $\Tt$, but at position $z$.
Finally, it moves back to the original position $y$.
For the latter step, we will exploit the fact that $R$ is bounded.

Now, for the details, we construct from $\gamma$ 
a finite monoid $(M,\cdot)$, 
a monoid morphism $h:(\Sigma\times\Sigma'\times\bbB\times\bbB)^*\rightarrow M$, and
some subsets $F_\otype$ of $M_\otype$, for each output-type $\otype\in\Gamma$, 
such that $(u,x,y)\sat\gamma(\otype)$ if and only if 
$h(u_{x,y})\in F_\otype$, where $u_{x,y}$ is the encoding on $u$
of the positions $x$ and $y$, namely,
$u_{x,y}(i)=\big((u(i),b_{i=x},b_{i=y})\big)$ for all $1\le i\le|u|$,
and $b_{i=x}$ (resp.~$b_{i=y}$) is either $1$ or $0$ depending on 
whether $i=x$ (resp.~$i=y$) or not.
Similarly, we denote by $u_{\emptyset,\emptyset}$ 
the encoding on $u$ of two empty monadic predicates. 
For all $1\le i\le j\le|u|$, we then define
$\ell_i=h(u_{\emptyset,\emptyset}[1,i-1])$, 
$r_j=h(u_{\emptyset,\emptyset}[j+1,|u|])$, and 
$m_{i,j}=h(u_{i,j}[i,j])$.
We observe that 
\[
  (u,y,z)\sat\gamma(\otype)
  \qquad\text{iff}\qquad
  \begin{cases}
    \ell_y\cdot m_{y,z}\cdot r_z\in F_\otype  & \text{if $y\le z$} \\
    \ell_z\cdot m_{z,y}\cdot r_y\in F_\otype  & \text{if $y>z$}.
  \end{cases}    
\]
The elements $\ell_i$ and $r_i$ associated with each position $i$ 
of the input $u$ are functionally determined by $u$. In particular
the word $\ell_1\dots\ell_{|u|}$ (resp.~$r_1\dots r_{|u|}$) can be 
seen as the run of a deterministic (resp.~co-deterministic) 
automaton on $u$.
Without loss of generality, we can assume
that the values $\ell_i$ and $r_i$ are readily available as 
annotations of the input at position $i$, and checked by means 
of a suitable refinement $\Aa'$ of the NFA $\Aa$.

We now describe how $\Tt'$ simulates a transition of $\Tt$,
say $q \act{a|b} q'$, that originates at a position $y$ and produces the letter $b$
(the simulation of a transition with empty output is straightforward).
The transducer $\Tt'$ stores in its control state the transition rule to
be simulated and the monoid element $\ell_y$ associated with the current position $y$
(the source). 
It then guesses whether the displaced origin $z$ (i.e.~the target) 
is to the left or to the right of $y$.
We only consider the case where $z\ge y$ (the case $z<y$ is symmetric).
In this case $\Tt'$ starts moving to the right, until it 
reaches some position $z\ge y$ such that $(u,y,z)\sat\gamma(b)$ 
(as we explained earlier, this condition is equivalent to 
checking that $\ell_y\cdot m_{y,z}\cdot r_z\in F_b$). 
Once a target $z$ is reached, $\Tt'$ produces the same
output $b$ as the original transition, 
and begins a new phase for backtracking to the source $y$.
During this phase, the transducer will maintain the previous monoid elements
$\ell_y$, $m_{y,z}$, and, while moving leftward, compute $m_{z',z}$ for all
$z'\le z$. We claim that there is a unique $z'$ such that
$\ell_{z'}=\ell_y$ and $m_{z',z}=m_{y,z}$, and hence such $z'$ 
must coincide with the source $y$. Indeed, if this were not the case,
we could pump the factor of the input between the correct source $y$ 
and some $z'\neq y$, showing that the MSO resynchronizer $R_\gamma$ 
is not bounded.
Based on this, the transducer $\Tt'$ can move back to the correct source $y$, 
from which it can then simulate the change of control state from $q$ 
to $q'$ and move to the appropriate next position.
Any run of $\Tt'$ that simulates a run of $\Tt$ on input $u$, 
as described above, results in producing the same output $v$ as $\Tt$, 
but with the origin mapping modified from $i\mapsto y_i$ to 
$i\mapsto z_i$, for all $1\le i\le |v|$ and for some $1\le y_i,z_i\le |u|$ 
such that $(u,y_i,z_i)\sat\gamma(v(i))$.
In other words, we have 
$\sem{\Tt'}_o = 
 \{ \oelement' \,:\, (\oelement,\oelement')\in\sem{R},\, \oelement\in\sem{\Tt}_o \}$.
 
Now, assume that the formulas of $\gamma$ 
(and hence the monoid $(M,\cdot)$) are fixed,
and let $n$ be the number of states of $\Tt$.
A close inspection to the above constructions shows 
that $\Tt'$ has a number of state polynomial in $n$.
Moreover, if $\Tt$ has \PSPACE-constructible transitions 
w.r.t.~$n$ and $\Aa$ is \PSPACE-constructible w.r.t.~$n$,
then $\Tt'$ and $\Aa'$ have similar properties.

\smallskip
We finally consider the case of 
$R_\delta=(\true,\true,\gamma_{\mathit{id}},\delta)$,
which restricts the origin mapping in such a way that pairs of 
origins $z,z'$ associated with consecutive outputs positions $x$ and $x+1$ 
satisfy the formula $\gamma(\otype)(z,z')$, where $\otype$ is the ouput-type of $x$
(in fact, the output symbol at $x$).
To deal with this case, it is convenient to first normalize our
transducer $\Tt$ in such a way that there are no lazy U-turns.
This normalization step was described in Section \ref{sec:arbitrary}
and results in a two-way transducer $\busy(\Tt_U)$ over an expanded 
input alphabet $\Sigma\times 2^{\Qleft\times\Qright}\times 2^{\Qright\times\Qleft}$
such that, when receiving an input $u$ correctly annotated with sets of $U$-pairs, 
produces the same outputs as $\Tt$ on $u$,
but does so while avoiding lazy U-turns. 
In other words, between any two consecutive non-empty outputs, 
the transducer $\busy(\Tt_U)$ will move along the shortest 
possible input path (that is either left-to-right or right-to-left).
We recall that $n$ is the number of states of $\Tt$, and that
$\busy(\Tt_U)$ has the same number $n$ of states and 
\PSPACE-constructible transitions w.r.t.~$n$. 
Moreover, to work correctly, $\busy(\Tt_U)$ needs to receive
an input that is correctly annotated with sets of $U$-turns:
such an input can be checked by the \PSPACE-constructible NFA 
$\Uu$ described again in Section \ref{sec:arbitrary}.

The next step is to construct from $\busy(\Tt_U)$ a
new transducer $\Tt'$ that restricts the possible outputs in 
an appropriate way, based on the formulas $\delta(\otype,\otype')(z,z')$.
We begin by following the same approach as before: we construct
a finite monoid $(M,\cdot)$, a monoid morphism
$h:(\Sigma\times\Sigma'\times\bbB\times\bbB)^*\rightarrow M$,
and some subset $F_{\otype,\otype'}$ of $M$, 
for each pair of output-types $\otype,\otype'\in\Gamma$,
such that $(u,z,z')\sat\delta(\otype,\otype')$
if and only if $h(u_{z,z'}))\in F_{\otype,\otype'}$. 
When moving from a position $z$ to another position $z'$,
we compute the monoid element
$h(u_{z,z'})$, which can be written as either
$\ell_z\cdot m_{z,z'}\cdot r_{z'}$ or $\ell_{z'}\cdot m_{z',z}\cdot r_z$,
depending on whether $z\le z'$ or $z>z'$, where 
$\ell_z=h(u_{\emptyset,\emptyset}[1,z-1])$,
$r_{z'}=h(u_{\emptyset,\emptyset}[z'+1,|u|])$, and
$m_{z,z'}=h(u_{z,z'}[z,z'])$.
As usual, the elements $\ell_z$ and $r_{z'}$ are assumed to be 
available as explicit annotations of the input (this requires 
expanding further the input alphabet and refining the NFA $\Aa$).
The element $m_{z,z'}$, on the other hand, can be computed by the 
transducer while moving along the shortest path from $z$ to $z'$.

We now specify how $\Tt'$ simulates a run of $\busy(\Tt_U)$ (and hence of $\Tt$),
while restricting the set of possible outputs.
As explained above, $\Tt'$ maintains in its control state
the monoid elements $\ell_z$, $m_{z,z'}$, and $r_z$, where 
$z$ is the origin of the last non-empty output and $z'$ is 
the current input position (for simplicity, here we assume
that $z\le z'$, otherwise we swap the roles of $z$ and $z'$).
In addition, it also maintains the most recently produced output 
letter, denoted $b$.
Now, if $\busy(\Tt_U)$ takes a transition from the current 
position $z'$ to a position $z''$, say with $z''\ge z$, 
and outputs the language $L\subseteq\Gamma^*$, then $\Tt'$ 
behaves as follows:
\begin{itemize}
  \item it non-deterministically chooses 
        a language among $L_\e$ and $L_{b'}$, for any $b'\in\G$, where
\[
\begin{array}{rcl}
  L_\e &=& L\cap \{\e\} \\[1ex]
  L_{b'}  &=& \left\{ v \in L \cap (\Sigma^*\cdot \{b'\} \::\: 
                     \begin{aligned}
                       & \ell_{z}\cdot m_{z,z'}\cdot r_{z'} \in F_{b,v(1)} 
                         \quad \text{if $b$ is last output letter} \\
                       & \ell_{z'}\cdot m_{z',z'}\cdot r_{z'} \in F_{v(i),v(i+1)} 
                         \quad \forall 1\le i<|v| 
                     \end{aligned} ~ \right\}
\end{array}
\]
  \item if it chose to output $L_\e$, then, assuming $L_\e\neq\emptyset$,
        it simulates
        the change of control state of the  
        transition of $\busy(\Tt_U)$, moves from $z'$ 
        to $z''$, and updates the stored monoid element 
        from $m_{z,z'}$ to $m_{z,z''}=m_{z,z'}\cdot h(a)$,
        where $a$ is the input letter at position $z''$;
  \item if it chose to output $L_{b'}$, then, assuming $L_b\neq\emptyset$,
        it simulates
        the change of control state of the transition of 
        $\busy(\Tt_U)$, moves from $z'$ to $z''$, 
        updates the stored monoid element from 
        $m_{z,z'}$ to $m_{z'',z''}=h(\e)$, 
        and finally changes the most recent output letter
        to $b'$.
\end{itemize}
In particular, the above behaviour guarantees the following two properties:
\begin{enumerate}
  \item $\delta(v(x),v(x+1))(z',z')$ holds for all pairs of consecutive output positions
         $x$ and $x+1$ that are produced at the same input position $z'$.
  \item $\delta(b,v(x+1))(z,z')$ holds for all pairs of consecutive output positions
        $x$ and $x+1$ that have origins respectively in $z$ 
        ---i.e.~the last input position where a non-empty output 
           ending with $b=v(x)$ was emitted---
        and $z'$ ---i.e.~the current input position, where a new
        non-empty output is going to be emitted.
\end{enumerate}
More generally, one can verify that the synchronized pairs in $\sem{\Tt'}_o$ 
are precisely those of the form $\oelement=(u,\nu)\in\sem{\Tt}_o$ such that
for all positions $x$ in $\Out(\oelement)$ labelled by $\otype$, 
if $\Orig(\oelement')(x)=z$ and $\Orig(\oelement')(x+1)=z'$, 
then $(\In(\oelement),z,z')\sat\delta(\otype)$.
This implies that 
$\sem{\Tt'}_o\prj[\Aa]{\Sigma} = 
 \{ \oelement' \,:\, \oelement\in\sem{\Tt}_o\prj[\Aa]{\Sigma},\, 
    (\oelement',\oelement)\in\sem{R_\delta} \}$.

With the above constructions, if $R$ is fixed and 
if $\Tt$ has $n$ states and \PSPACE-constructible transitions 
w.r.t.~$n$, then similarly $\Tt'$ has a number of states 
polynomial in $n$ and \PSPACE-constructible transitions w.r.t.~$n$.
Finally, a \PSPACE-constructible NFA $\Aa'$ can be obtained from a direct product
of the NFA $\Aa$, $\Uu$, and a suitable NFA for checking inputs annotated with
the monoids elements $\ell_z,r_z$.
\end{proof}

The above result is used to reduce the 
containment problem modulo a bounded MSO resynchronizer 
$R$ to a containment problem with origins 
(relativized to correctly annotated inputs).
That is, if $\Tt_1,\Tt_2$ are transducers with input alphabet $\Sigma$,
and $\Tt'_2,\Aa'$ are over the input alphabet $\Sigma\times\Sigma'$ and
constructed from $\Tt_2$ using Theorem~\ref{thm:resync}, 
then
\[
  \Tt_1 \subseteq_R \Tt_2
  \qquad\text{iff}\qquad
  \sem{\Tt_1}_o \:\subseteq\: \sem{\Tt_2'}_o \prj[\Aa']{\Sigma}
  \qquad\text{iff}\qquad
  \sem{\Tt_1}_o\lift{\Sigma\times\Sigma'}  ~\cap~ R' \:\subseteq\: \sem{\Tt_2'}_o ~\cap~ R'
\]
where $R'=\sem{\Aa'}\times(\G\times\Nat)^*$ and 
$\lift{\Sigma'}$ is the inverse of the input-projection operation, 
i.e.~$\sem{\Tt}_o\lift{\Sigma\times\Sigma'} 
  = \big\{ (u\otimes u',\nu) \in (\S\times\S')^*\times(\G\times\Nat)^* \::\:
           u\otimes u'\in\sem{\Aa}, ~ (u,\nu)\in\sem{\Tt}_o \big\}$.
We also recall from Section \ref{sec:equiv} that the latter containment 
reduces to emptiness of a \PSPACE-constructible NFA, which can 
then be decided in \PSPACE w.r.t.~the sizes of $\Tt_1$ and $\Tt_2$. 
We thus conclude with the following result:

\begin{corollary}\label{thm:containment-modulo}
The problem of deciding whether $\Tt_1\subseteq_R \Tt_2$,
for any pair of transducers $\Tt_1,\Tt_2$ and for a fixed
bounded MSO resynchronizer $R$, is \PSPACE-complete.
\end{corollary}

 \section{Conclusions}\label{sec:conclusions}

We studied the equivalence and containment problems for non-deterministic, 
two-way word transducers in the origin semantics, and proved that the problems
are decidable in \PSPACE, which is the lowest complexity one could expect
given that equivalence and containment of NFA are already \PSPACE-hard.
This result can be contrasted with the undecidability of equivalence
of non-deterministic, one-way word transducers in the classical
semantics.

We have also considered a variant of containment 
up to  `distortions' of the origin, called containment 
modulo a resynchronization $R$, and denoted $\subseteq_R$. 
We identified a broad class of resynchronizations, definable 
in MSO, and established decidability of the induced containment problem. 
In fact, we obtained an optimal fixed-parameter complexity result: 
testing a containment $\Tt_1\subseteq_R\Tt_2$ modulo a bounded MSO 
resynchronization $R$ is \PSPACE-complete in the size of the input 
transducers $\Tt_1,\Tt_2$, where the fixed parameter is the size
of the formulas used to describe $R$.

\medskip
Our logical definition of resynchronizations talks implicitly about
origin graphs.
Since we cannot encode the  origin semantics of arbitrary two-way
transducers by words, we have chosen to 
work directly with origin graphs  defined by two-way transducers. 
A classical way to define resynchronizations of origin graphs would be to use 
logical formalisms for graph transformations. 
Unfortunately, the classical MSO approach~\cite{Cou97,eng97} does not work
in our setting, since satisfiability of MSO over origin graphs of
two-way transducers
is already undecidable, which means that the definable resynchronizations
would not be realizable by two-way transducers.

Another possibility would be to use a decidable logic over origin
graphs, like the one introduced by Filiot et al.~in~\cite{dfl18lics}.
Their logic is not suited either for our purposes, since it allows
predicates of arbitrary arity defined using MSO over the input word.
The reason is that single head devices would not be able to move
between the tuples of positions related by those definable predicates,
and thus, in particular, we would not be able to guarantee that
the definable resynchronizations are realizable by two-way transducers.

Yet an alternative approach to the above problem consists in viewing
tagged outputs as data words,
and using transducers to transform data words. Durand-Gasselin and
Habermehl introduced in~\cite{dgh16fossacs} a framework for
transformations of  data words, where MSO transductions, deterministic
two-way transducers 
and deterministic streaming transducers have the same expressive
power. However, this approach would be unsatisfactory, because the
transformation does not take the input into account.

\medskip
We conjecture that the bounded MSO resynchronizers defined here strictly 
capture the rational resynchronizers introduced in~\cite{FiliotJLW16}. 
In particular, although MSO resynchronizers do not have the ability 
to talk about general origin graphs, they presumably can describe `regular' 
origin graphs of one-way transducers, i.e., graphs expressed by regular 
languages over sequences that alternate between the input and the output word.
We also recall that the MSO resynchronizer $R_{\mathit{1st-to-last}}$ 
from Example \ref{ex:resync}.5 contains only origin graphs of one-way 
transducers, but cannot be defined by a rational resynchronizer.

Another natural question that we would like to answer concerns 
compositionality: are bounded MSO resynchronizers closed under relational composition?
In other words, given two bounded MSO resynchronizers $R,R'$, is it possible to find (possibly effectively)
a bounded MSO resynchronizer $R''$ such that $\sem{R''}=\sem{R}\circ\sem{R'}$? 
\bibliographystyle{alpha}
\bibliography{biblio}

\end{document}